\documentclass[12pt]{article}
\usepackage{latexsym}

\newlength{\abstractwidth}
\input epsf

\renewcommand{\thanks}[1]{\footnote{#1}} % Use this for footnotes

\newcommand{\bea}{\begin{eqnarray}}
\newcommand{\eea}{\end{eqnarray}}
\newcommand{\be}{\begin{eqnarray*}}
\newcommand{\ee}{\end{eqnarray*}}
\newcommand{\<}{\langle}
\renewcommand{\>}{\rangle}

\usepackage{epsfig}

\makeatletter
\newcounter{fig}
\renewcommand\thefig{\arabic{fig}}

\def\fps@fig{tbp}
\def\ftype@fig{1}
\def\ext@fig{lof}
\def\fnum@fig{\figurename~\thefig}
\newenvironment{fig}
               {\@float{fig}}
               {\end@float}
\newenvironment{fig*}
               {\@dblfloat{fig}}
               {\end@dblfloat}

\def\A{{\cal A}}

\def\D{{\cal D}}
\def\F{{\cal F}}

\def\M{{\cal M}}
\def\N{{\cal N}}
\def\O{{\cal O}}

\def\X{{\cal X}}

\def\p{{\partial}}
\def\sdet{{\rm sdet}}
\def\tet{\vartheta}
\def\chiz{\chi _{\bar z}{}^+}
\def\chiw{\chi _{\bar w}{}^+}

\def\Im{{\rm Im}}

\def\half{{1\over 2}}
 
\def\z{{\bf z}}

\begin{document}
\baselineskip=16pt

\begin{flushright}
UCLA/02/TEP/28 \\
Columbia/Math/02
\end{flushright}

\bigskip

\begin{center}
{\Large \bf LECTURES ON TWO-LOOP SUPERSTRINGS}\footnote{Research
supported in part by National Science Foundation grants PHY-98-19686,
PHY-0140151, and DMS-98-00783.}

\bigskip\medskip

{\bf Hangzhou, Beijing 2002}

\bigskip\bigskip

{\large \bf Eric D'Hoker$^a$ and D.H. Phong$^b$} \\ 

\bigskip

$^a$ \sl Department of Physics and \\
\sl University of California, Los Angeles, CA 90095, USA \\
$^b$ \sl Department of Mathematics \\ 
\sl Columbia University, New York, NY 10027, USA

\end{center}

\bigskip\bigskip

\begin{abstract}
 
\medskip

In these lectures, recent progress on multiloop superstring perturbation
theory is reviewed. A construction from first principles is given 
for an unambiguous and slice-independent two-loop  superstring measure on
moduli space for even spin structure. A consistent choice of moduli,
invariant under local worldsheet supersymmetry is made in terms of the
super-period matrix. A variety of subtle new contributions arising from a
careful gauge fixing procedure are taken into account.

The superstring measure is computed explicitly in terms of genus two
theta-functions and reveals the importance of a new modular object of
weight 6. For given even spin structure, the measure exhibits a behavior
under degenerations of the worldsheet that is consistent with 
physical principles. The measure allows for a unique modular covariant GSO
projection. Under this GSO projection, the cosmological constant, the 1-,
2- and 3- point functions of massless supergravitons vanish pointwise
on moduli space. A certain disconnected part of the 4-point function is
shown to be given by a convergent integral on moduli space. A general
consistent formula is given for the  two-loop cosmological constant in
compactifications with central charge $c=15$ and with $\N=1$ worldsheet
supersymmetry. Finally, some comments are made on possible extensions of
this work to higher loop order.

\end{abstract}

\vfill\eject

%%%%%%%%%%%%%%%%%%%%%%%%%%%%%%%%%%%%%%%%%%%%%%%%%%%%%%%%%%%%%%%%%%%%%
%%%%%%%%%%%%%%%%%%%%%%%%%%%%%%%%%%%%%%%%%%%%%%%%%%%%%%%%%%%%%%%%%%%%%
\section{Introduction}
\setcounter{equation}{0}
%%%%%%%%%%%%%%%%%%%%%%%%%%%%%%%%%%%%%%%%%%%%%%%%%%%%%%%%%%%%%%%%%%%%%
%%%%%%%%%%%%%%%%%%%%%%%%%%%%%%%%%%%%%%%%%%%%%%%%%%%%%%%%%%%%%%%%%%%%%

In these Lectures, a review is presented of recent advances
\cite{I,II,III,IV} made in the conceptual understanding and concrete
calculation of two loop Type II and Heterotic superstring amplitudes. 
To tree and one loop level, the basic scattering amplitudes for the Type
II and Heterotic superstrings had been calculated in the foundational
papers \cite{gs82} and \cite{ghmr86}, where these string theories were
first constructed. To higher loop level, however, a reliable and concrete
formulation of even the simplest amplitudes (such as the zero point
function or cosmological constant) had remained unavailable until these
works. 

\medskip

The key complication at higher loop level, in the Ramond-Neveu-Schwarz
(RNS) formulation of superstring theory, is the emergence from the gauge
fixing procedure of odd-Grassmann-valued supermoduli \cite{fms,superm}.
Odd supermoduli are absent at tree level and at one-loop level with even
spin structures. There is an odd supermodulus to one-loop level with odd
spin structure, but its role is merely that of a bookkeeping device and is
dealt with easily \cite{dp88}. A graphical representation of the 4-point
function to tree, one and two loop levels is given in Fig.\ref{fig:1}.

\medskip

%%%%%%%%%%%%%%%%%%%%%%

\begin{fig}[b]
\centering
\epsfxsize=5in
\epsfysize=1.8in
\epsffile{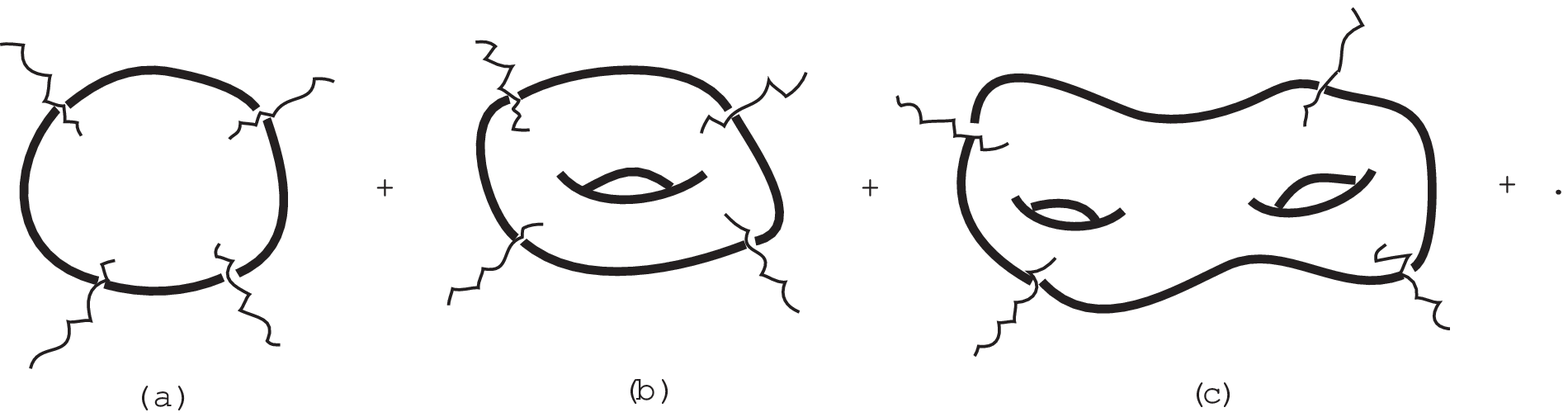}
\caption{String 4-point function to (a) tree-level, (b) one-loop level
and (c) two-loop level.}
\label{fig:1}
\end{fig}

%%%%%%%%%%%%%%%%%%%%%

A proposal based on worldsheet conformal field theory, BRST invariance
and the picture changing operator was made in \cite{fms}, in which it
was proposed to summarize the effects of odd supermoduli in terms of the
picture changing operator on a purely bosonic worldsheet specified by
bosonic moduli only. Although this approach is both natural and
appealing, a concrete calculation \cite{vv1} has demonstrated that to two
loop level, results are obtained that depend on the gauge slices
chosen, and are thus inconsistent. To remedy this situation, a general
procedure for generating correction terms and restoring slice
independence, based on Cech cohomology, was developed in \cite{v}. A
construction from first principles remained, however, out of reach and this scheme was not applied in practice.

\medskip

Considerable efforts were made by many authors to overcome the obstacles
identified in \cite{vv1} in terms of alternative prescriptions. These
drew from a variety of fundamental principles, such as modular invariance
\cite{mhns}, the lightcone gauge \cite{mand}, the global geometry of
Teichmuller space \cite{ams}, the unitary gauge \cite{lp,mz,iz}, the
operator formalism \cite{opf1,opf2}, group theoretic methods \cite{neveu},
factorization \cite{fact}, and algebraic geometry \cite{asg}. The basic
problem, however, that gauge-fixing required a local gauge slice and
that any consistent prescription must be independent of the choice of
such slice, remained unsolved. In fact, this state of affairs
raised the undesirable possibility that the definition of higher loop
superstring amplitudes could be inherently ambiguous \cite{ars1,ars2} and
that it may be necessary to consider other options, such as the
Fischler-Susskind mechanism \cite{ln}.

\medskip

In \cite{dp88} and \cite{dp89}, we had proposed that the difficulties
encountered in the earlier prescriptions were the result of improper
gauge-fixing procedures which did not respect worldsheet local
supersymmetry. As a point of departure, a superspace formulation of the
worldsheet \cite{superg1,superg2} was used. Superholomorphic anomalies,
generalizing those found for the bosonic string in \cite{div}, had
already been identified and their cancellation in the critical
superstring had already revealed the key role that would be played by
supermoduli space \cite{superanom}. Finally, a first principles gauge
fixing of the superstring amplitudes and chiral splitting were carried
out in \cite{dp88,dp89}, thereby producing a consistent formulation of
superstring amplitudes as integrals over supermoduli space. In
particular, this formulation was used in \cite{dp90} to show 
the perturbative unitarity of
the superstring amplitudes.

\medskip

Supermoduli space for higher genus surfaces is a delicate and complicated
object, and the ultimate goal in superstring perturbation theory is to
{\sl integrate out the odd supermoduli} and achieve a formulation in
terms of integrals over bosonic moduli only. It may be inferred from the
work of \cite{vv1} that the problems with the BRST picture changing
operator approach arise from an elimination of the odd supermoduli that
is inconsistent with local worldsheet supersymmetry. 
To take proper account of this supersymmetry, we had outlined in
\cite{dp88,dp89A} a new gauge-fixing procedure based on projecting
supergeometries onto their super period matrices instead of onto their
underlying bosonic geometries. Unlike the projection to the bosonic
geometries, the projection to the super period matrix is invariant under
local worldsheet supersymmetry. In the recent papers \cite{I,II,III,IV},
it is shown that this approach, applied to two loop level, solves all the 
problems encountered with the previous prescriptions.

\newpage

\subsection{Summary of results}

\hskip .255in $\bullet$ 
A gauge-fixed formula $d\mu[\delta](\Omega)$ for the contribution to the
superstring measure of each even spin structure $\delta$, which is {\sl
independent of the choice of gauge slice} was constructed from first
principles in \cite{II}. The ambiguities plaguing the earlier 
prescriptions have now disappeared, as was shown in \cite{II,III}.

\vskip .1in

$\bullet$ 
In \cite{IV}, the chiral measure $d\mu[\delta](\Omega)$ is evaluated
explicitly in terms of the genus two $\tet$-constants, and a new modular
object, $\Xi _6 [\delta](\Omega)$ emerges naturally from the
construction. For each $\delta$, $d\mu[\delta](\Omega)$ transforms
covariantly under modular transformations. There is a unique assignment
of relative phases $\eta _{\delta}$ so that $\sum_{\delta}\eta
_{\delta} \ d\mu[\delta](\Omega)$ is a modular form, and hence a unique
way of implementing the Gliozzi-Scherk-Olive (GSO) projection \cite{sw}.

\vskip .1in

$\bullet$ The superstring measure, when summed over all $\delta$, and
therefore also the cosmological constant, vanishes {\sl point by point} on
moduli space \cite{IV}. In establishing this property, use is made of a
2-loop generalization of the Jacobi identity which, remarkably, is not a
consequence of the genus 2 Riemann identities only. Instead, it is
equivalent to the identity, special to genus 2, that any modular form of
weight 8 must be proportional to the  square of the unique modular form
of weight 4.

\vskip .1in

$\bullet$ Similarly, the 1-, 2- and 3-point functions for the scattering
of the supergraviton multiplet vanish by a variety of novel identities. 
This result was announced in \cite{I}; a detailed proof will appear in
\cite{V}.

\vskip .1in

$\bullet$ The 4-point function may be evaluated explicitly in terms of
$\tet$-functions and modular forms. For a certain disconnected part of the
4-point function, explicit formulas are presented here; they are
manifestly finite, in the regime of purely imaginary Mandelstam
variables. As is well known \cite{dp94}, the other regimes are  
accessible only after proper analytic continuation in the external
momenta. The connected part and the full 4-point function will appear in
a forthcoming publication
\cite{VI}.

\vskip .1in

$\bullet$ Finally, we provide a simple slice independent formula for the
even spin structure superstring measure and cosmological constant for
general compactifications with matter central charge $c=15$ and
$\N=1$ worldsheet supersymmetry. This formula was announced in \cite{I};
it will be applied to the case of ${\bf Z}_2$ orbifolds in a forthcoming
publication \cite{adp}.

\vfill\eject

%%%%%%%%%%%%%%%%%%%%%%%%%%%%%%%%%%%%%%%%%%%%%%%%%%%%%%%%%%%%%%%%%%%%%
%%%%%%%%%%%%%%%%%%%%%%%%%%%%%%%%%%%%%%%%%%%%%%%%%%%%%%%%%%%%%%%%%%%%%
\section{The Ramond-Neveu-Schwarz Formulation}
\setcounter{equation}{0}
%%%%%%%%%%%%%%%%%%%%%%%%%%%%%%%%%%%%%%%%%%%%%%%%%%%%%%%%%%%%%%%%%%%%%
%%%%%%%%%%%%%%%%%%%%%%%%%%%%%%%%%%%%%%%%%%%%%%%%%%%%%%%%%%%%%%%%%%%%%

In the RNS formulation \cite{rns} of superstring theory, the fundamental
string degrees of freedom are the bosonic position  $x^\mu$ and the
fermionic counterpart $\psi ^\mu$. Both are fields on the worldsheet
$\Sigma$ and transform as vectors under the Lorentz transformations of
flat Minkowski space-time. The formulation also appeals to the worldsheet
metric $g_{mn}$ and gravitino $\chi_m$ fields, which are non-dynamical.
The starting point for the formulation of scattering amplitudes is the
worldsheet supergravity action of \cite{bdh}, given by
\bea 
\label{action}
I_m & = & {1 \over 4 \pi} \int _\Sigma d^2z \sqrt{g}  \biggl (
\half g^{mn} \p_m x^\mu  \p_n x^\mu + \psi ^\mu \gamma ^m \p_m \psi ^\mu 
\nonumber \\ && \hskip .8in
- \psi ^\mu \gamma ^n \gamma ^m \chi _n \p_m x^\mu 
- {1 \over 4} \psi ^\mu \gamma ^n \gamma ^m 
\chi _n (\chi _m \psi ^\mu) \biggr  )
\eea
The action is constructed so as to be invariant under diffeomorphisms,
local $\N=1$ supersymmetry, Weyl and super Weyl transformations of the
worldsheet. In view of the key role played by local supersymmetry, it is
convenient to reformulate the action in terms of a matter superfield
$X^\mu$ and a supergeometry specified by a local frame $E_M{}^A$
and local $U(1)$ connection superfield $\Omega _M$
\cite{superg1,superg2,superanom,dp88}. For a brief summary see Appendix~B
of \cite{II}. The relation between component and superfields is,
($A$ and $F^\mu$ are auxiliary fields)
\bea
X^\mu & \equiv & x^\mu + \theta \psi ^\mu _+ + \bar \theta \psi _- ^\mu +
i \theta \bar \theta F^\mu 
\nonumber \\
E_m {}^a & \equiv & e_m {}^a + \theta \gamma ^a \chi _m - {i \over 2}
\theta \bar \theta A e_m {}^a
\eea
In terms of these superfields, the worldsheet action (\ref{action}) takes
the simple form,
\bea
I_m 
={1\over 4\pi} \int _\Sigma d^{2|2}\z \, E\, {\cal D}_+X^{\mu}{\cal
D}_-X^{\mu}
\qquad \quad
E \equiv \sdet E_M{}^A
\eea
where $\D_\pm$ are supercovariant derivatives, whose precise form may be
found in Appendix B of \cite{II} but will not be needed here. 

\medskip

The starting point for the scattering amplitudes is the Polyakov
formulation of string perturbation theory \cite{polyakov}, in which a
summation is performed over all surfaces (including their topologies,
specified by the number of handles $h$) and all fields on the surface,
\bea
{\bf A}_\O = \sum _{h=0}^\infty  \int {D(E\Omega) \ \delta (T)
\over {\rm Vol \ (Symm)}} \int DX^\mu \ \O \ e^{-I_m}
\eea 
The operator $\O$ stands for the insertion of any set of physical state
vertex operators, whose construction in superspace was given in
\cite{dpvertex}. In the critical dimension, 10, the quantum string is
invariant under the full set of classical symmetries,
\bea
\label{symmetries}
{\rm Symm} = {\rm sDiff}(\Sigma) \times {\rm sWeyl}(\Sigma) \times {\rm
sU}(1) (\Sigma)
\eea
which must be factored out. Finally, $\delta (T)$ indicates that the
torsion constraints of the $\N=1$ supergeometry are to be enforced.

\vfill\eject

%%%%%%%%%%%%%%%%%%%%%%%%%%%%%%%%%%%%%%%%%%%%%%%%%%%%%%%%%%%%%%%%%%%%%
%%%%%%%%%%%%%%%%%%%%%%%%%%%%%%%%%%%%%%%%%%%%%%%%%%%%%%%%%%%%%%%%%%%%%
\section{Reliable Superspace Gauge Fixing}
\setcounter{equation}{0}
%%%%%%%%%%%%%%%%%%%%%%%%%%%%%%%%%%%%%%%%%%%%%%%%%%%%%%%%%%%%%%%%%%%%%
%%%%%%%%%%%%%%%%%%%%%%%%%%%%%%%%%%%%%%%%%%%%%%%%%%%%%%%%%%%%%%%%%%%%%

As was shown in \cite{dp88}, a reliable gauge fixing procedure may be
derived from first principles by reducing the integral over all
supergeometries to a finite-dimensional integral over the quotient of all
supergeometries by all the local symmetries of (\ref{symmetries}). This
quotient is referred to as {\sl supermoduli space}; its dimensions are as
follows,
\bea
s\M_h  & \equiv &  \{E_M{}^A, \Omega _M + {\rm torsion \ constraints} \}
\big  / {\rm sDiff } \times {\rm sWeyl} \times {sU(1)}
\nonumber \\ && \nonumber \\
{\rm dim} (s\M _h) & = & 
\cases{
(0|0) & $h=0$ \cr 
(1|0)_e \ {\rm or} \ (1|1)_o & $h=1$ \cr
(3h-3|2h-2) & $h\geq 2$ \cr} 
\eea
The subscripts $e$ and $o$ refer to the cases of 
even and odd spin structures respectively.

\medskip

Being a quotient, supermoduli space does not admit a canonical
parametrization, and one is led to choosing a local slice ${\cal S}$ of
the same dimension as $s\M_h$, and transverse to the orbits of the
symmetry group  ${\rm sDiff}(\Sigma) \times {\rm sWeyl}(\Sigma) \times
{\rm sU}(1) (\Sigma)$. Specializing to $h\geq 2$, we parametrize ${\cal
S}$ by $m^A = (m^a|\zeta ^\alpha)$, $a=1, \cdots ,3h-3$ even and $\alpha
=1,\cdots ,2h-2$ odd supermoduli. A gauge fixed formulation in terms of a
slice representing supermoduli space $s\M_h$ was proposed in \cite{vv1}
and derived from first principles in \cite{dp88}. It involves ghost
superfields $B$ and
$C$ (as well as their complex conjugates) which are related to the
customary ghost fields $b$ and $c$ and superghost fields $\beta$ and
$\gamma$ by
\bea
B & \equiv & \beta + \theta b + {\rm auxiliary \ fields} 
\nonumber \\
C& \equiv & c + \theta \gamma + {\rm auxiliary \ fields} 
\eea
The gauge fixed expression for the amplitudes is given by
\bea
\label{nonchiral}
{\bf A} _\O = \int_{s\M} \! |dm^A|^2 \int \! \! D(X B C) \ 
\big | \prod _A \delta (\< H_A |B\> ) \big |^2 \ \O \ e^{-I} 
\eea
The combined matter and ghost action is given by
\bea
I  \equiv  {1\over 2 \pi} \int _\Sigma \! d^{2|2}\z E \ 
\biggl ( \half \D_+ X^\mu \D_- X_\mu +
 B \D_- C + \bar B \D _+ \bar C \biggr )
\eea
The super-Beltrami differential, defined by
\bea
\label{superbeltrami}
( H_A ) _- {}^z \equiv (-)^{A(M+1)} E_- {}^M {\p E_M{}^z \over \p m^A}
= \bar \theta (\mu_A - \theta \chi _A) \bigg | _{\rm WZ}
\eea
represents the tangent vectors to the slice ${\cal S}$.

\vfill\eject

%%%%%%%%%%%%%%%%%%%%%%%%%%%%%%%%%%%%%%%%%%%%%%%%%%%%%%%%%%%%%%%%%%%%%
%%%%%%%%%%%%%%%%%%%%%%%%%%%%%%%%%%%%%%%%%%%%%%%%%%%%%%%%%%%%%%%%%%%%%
\section{Chiral Splitting}
\setcounter{equation}{0}
%%%%%%%%%%%%%%%%%%%%%%%%%%%%%%%%%%%%%%%%%%%%%%%%%%%%%%%%%%%%%%%%%%%%%
%%%%%%%%%%%%%%%%%%%%%%%%%%%%%%%%%%%%%%%%%%%%%%%%%%%%%%%%%%%%%%%%%%%%%

The formulation of the amplitudes given in (\ref{nonchiral}) is such that
left moving (or holomorphic) and right moving (or anti-holomorphic)
degrees of freedom are related to one another by complex conjugation.
This result emerges naturally when the starting point of string theory is
in terms of a summation over actual surfaces with Euclidean signature
worldsheet metrics. In some deep sense, the original theory was rather in
terms of a Minkowskian worldsheet where the fermions of left
and right chiralities are independent of one another. This independence
is a crucial ingredient in the very definition of both the Type II and
Heterotic string theories, since left and right chiralities are assigned
independent spin structures and a GSO projection must be carried out
independently on left and right chirality degrees of freedom.

\medskip

To recover the independence of left and right chiralities on a worldsheet
with an Euclidean signature metric, a process of splitting the
chiralities from one another must be applied. A glance at the worldsheet
action (\ref{action}) immediately reveals that this process appears to
have some basic obstructions; the quartic term in fermions couples left
and right chiralities to one another and the zero mode of the scalar
field $x^\mu$ cannot be split. Similar obstructions appear when vertex
operators are included. Nonetheless, the splitting is possible within each
conformal block, labeled here by the internal loop momenta, and the
chiralities may be identified with holomorphic and anti-holomorphic
dependence on supermoduli \cite{superanom,dp88,dp89}.  

\medskip

The chiral splitting procedure may be summarized in terms of a set of
effective rules \cite{dp89,dp90}, which we now spell out. Chiral
splitting may be achieved for each conformal block, which is
labelled uniquely by a set of $h$ independent internal loop momenta
$p_I^{\mu}$, $ I=1,\cdots,h$. It will be convenient to choose a basis for
the first homology of the surface in terms of canonical $A_I$ and $B_I$,
$I=1,\cdots ,h$ cycles, such as depicted in Fig.\ref{fig:2} for genus 2.

%%%%%%%%%%%%%%%%%%%%%%

\begin{fig}[htp]
\centering
\epsfxsize=4in
\epsfysize=1.7in
\epsffile{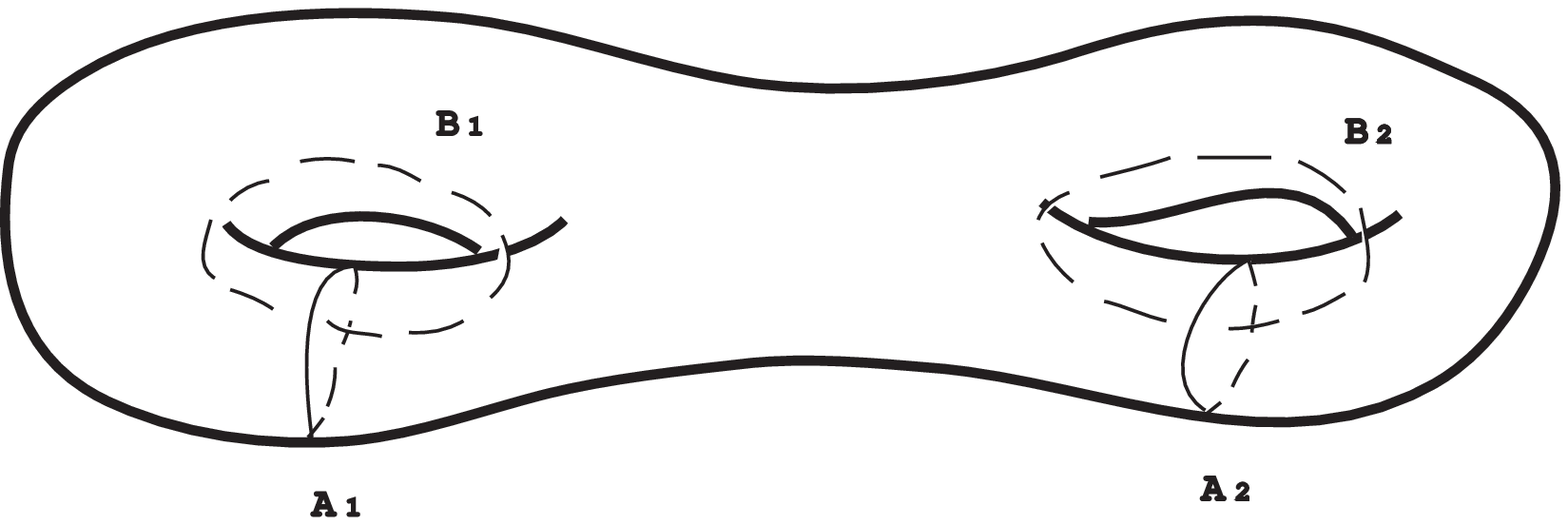}
\caption{Choice of Canonical homology basis for genus 2}
\label{fig:2}
\end{fig}

%%%%%%%%%%%%%%%%%%%%%

The independent internal loop momenta $p_I^{\mu}$, $ I=1,\cdots,h$ may
then be viewed as the momenta traversing the cycles $A_I$.
The following effective prescription for the scalar superfield correlation
functions emerges,
\bea
\label{chispl}
\<\prod_{i=1}^NV_i(k_i, \epsilon _i )\>_{X^{\mu}}
=
\int dp_I^{\mu}\ \bigg | \bigg \< \prod_{i=1}^N
V_{i}^{chi}(k_i, \epsilon _i;p_I^{\mu})
\bigg \>_+ \bigg | ^2
\eea
Here, $\<\cdots \>_+$ denotes the fact that the effective rules for the
contractions of the vertex operators $V_{i}^{chi}(k_i, \epsilon _i
;p_I^{\mu})$ are used, as given in Table 1.

\medskip

%%%%%%%%%%%%%%%%%%%%%%%%%%%%%%%%%%%%%%%%%%%%%%%%%%%%%%%%%%%%%%%%%%%%%%
%%%%%%%%%%%%%%%%%%%%%%%%%%%%%%%%%%%%%%%%%%%%%%%%%%%%%%%%%%%%%%%%%%%%%%
\begin{table}[htb]
\begin{center}
\begin{tabular}{|c||c|c|} \hline 
& {\rm Original} & {\rm Effective Chiral} 
                \\ \hline \hline
              {\rm Bosons}  
            & $x^{\mu}(z)$ 
            & $x_+^{\mu}(z)$        
             
 \\ \hline
              Fermions  
            & $\psi_+ ^\mu (z)$ 
            & $\psi_+ ^\mu (z)$
\\ \hline
 Internal Loop momenta
& None
& ${\rm exp}(p_I^{\mu}\oint_{B_I}dz\partial_zx^{\mu} _+)$       
            
 \\ \hline
              $x$-propagator  
            & $\<x^{\mu}(z)x^{\nu}(w)\>$ 
            & $-\delta^{\mu\nu}{\rm ln}\,E(z,w)$        
 \\ \hline
              $\psi_+$-propagator  
            & $\<\psi_+ ^\mu (z)\psi_+^{\nu}(w)\>$ 
            & $- \delta^{\mu\nu}S_{\delta}(z,w)$        
 \\ \hline
              Covariant Derivatives  
            & ${\cal D}_+$ 
            & $\partial_{\theta}+\theta\partial_z$        
 
 \\ \hline
\end{tabular}
\end{center}
\caption{Effective Rules for Chiral Splitting}
\label{table:1}
\end{table}
%%%%%%%%%%%%%%%%%%%%%%%%%%%%%%%%%%%%%%%%%%%%%%%%%%%%%%%%%%%%%%%%%%%%%%
%%%%%%%%%%%%%%%%%%%%%%%%%%%%%%%%%%%%%%%%%%%%%%%%%%%%%%%%%%%%%%%%%%%%%%

\medskip

In this table, $E(z,w)$ is the prime form, and $S_\delta (z,w)$ is
the Szeg\"o kernel. The point of the effective rules is that they only
involve meromorphic notions, unlike the $x$-propagator
$\<x^{\mu}(z)x^{\nu}(w)\>$ which is given by the scalar Green's function
$\delta ^{\mu\nu}G(z,w)$. The superghost correlation functions are
manifestly split. We obtain the following formula,
\bea
\label{smeaschi}
{\bf A}_\O [\delta] 
=
\int |\prod_A dm^A|^2
\int dp_I^{\mu}\
\bigg | e^{i\pi p_I^{\mu}\hat\Omega_{IJ}p_J^{\mu}}\,
{\cal A} _\O [\delta]\, \bigg |^2\, 
\eea
where ${\cal A} _\O[\delta]$ is the following effective chiral correlator
\bea
\label{smea}
{\cal A} _\O [\delta] 
=
\bigg \< 
\prod_A\delta(\<H_A|B\>) \ \O_+ \ 
\exp \biggl \{ \int _\Sigma \! {d^2\! z \over 2 \pi} \chi_{\bar z}{}^+
S(z) \biggr  \}\bigg \> _+
\eea
and $S(z)$ is the total supercurrent
\bea
S(z)
=-{1\over 2}\psi_+^{\mu}\p_zx_+^{\mu}
+{1\over 2}b\gamma-{3\over 2}\beta\p_zc
-(\p_z\beta)c,
\eea
Here, $\hat \Omega_{IJ}$ is the super period matrix, defined by
\cite{dp88,dp89} for any genus. For genus 2, its expression simplifies
considerably and is given by
\bea
\label{sper}
\hat\Omega_{IJ}
=
\Omega_{IJ}-{i\over 8\pi}\int _\Sigma \! d^2 \! z \int _\Sigma \! d^2 \! w
\ \omega_I(z) \chiz  S_{\delta}(z,w) \chiw \omega_J(w)
\eea
Here $\Omega_{IJ}$ is the period matrix corresponding to the
complex structure of the metric $g_{mn}$.  The $\omega_I(z)$ span a basis
of holomorphic Abelian differentials dual to the $A_I$-cycles, so that
\bea
\oint _{A_I} \omega _J = \delta _{IJ}
\qquad \qquad 
\oint _{B_I} \omega _J = \Omega _{IJ}
\eea
The period matrix may also be obtained in an intrinsic way from the
superholomorphic 1/2 forms $\hat \omega _I$, which are the super analogs
of the ordinary holomorphic Abelian 1-forms $\omega _I$. Given again the
choice of canonical homology cycles as depicted in Fig.\ref{fig:2}, the
$\hat \omega _I$ may be canonically normalized on $A_I$-cycles and yield
the super period matrix when integrated over $B_I$ cycles, 
\bea
\D _- \hat \omega _I =0 \hskip 1in
\oint _{A_I} \hat \omega _J =\delta _{IJ}
\qquad 
\oint _{B_I} \hat \omega _J = \hat \Omega _{IJ}
\eea
in complete analogy with the oridnary Abelian differentials.

\medskip

The chirally split expression (\ref{smeaschi}) and (\ref{smea}) is our
first significant departure \cite{dp88,dp89} from the proposals of other
authors in the late 1980's, in that it is the super period matrix
$\hat\Omega_{IJ}$ which appears as covariance of the internal loop
momenta $p_I^{\mu}$, and not the ordinary period matrix $\Omega_{IJ}$.
Therefore, a correct chiral splitting points to the super period matrix
$\hat\Omega_{IJ}$ as the proper locally supersymmetric moduli for
gauge-fixing.

\medskip

The actual amplitudes for the Type II and Heterotic superstrings are 
obtained by assembling the contributions from left and right movers
endowed with the same period matrix and internal momenta, but
with independent spin structures (or winding sectors for the
bosonic formulation of the right moving part of the heterotic string).
The correct amplitudes are then given by
\bea
{\bf A} _{II \O} & = & \int dp^\mu _I \sum _{\delta , \bar \delta}
\eta_{\delta, \bar \delta}
\int _{s\M _h} |dm^A|^2 |\exp \{ i \pi p^\mu _I \hat \Omega _{IJ} p^\mu
_J\}| {\cal A}_\O [\delta ](\hat \Omega) \bar {\cal A}_\O [\bar \delta
](\hat \Omega ^*) 
\nonumber \\
{\bf A} _{H \O} & = & \int dp^\mu _I \sum _\delta  \eta _\delta
\int _{s\M _h} dm^A \int _{\M_h} d\bar m^a |\exp \{ i \pi p^\mu _I \hat
\Omega _{IJ} p^\mu _J\} {\cal A}_\O [\delta ](\hat \Omega) \bar {\cal
B}_\O ( \Omega ^*) 
\eea
where ${\cal B}_\O$ stands for the chiral half of the 26-dimensional
heterotic string, compactified on a self-dual even lattice, suitable for
the heterotic string. Here, $\M_h$ stands for the bosonic moduli space
of Riemann surfaces at genus $h$. Note that the period matrix
$\Omega_{IJ}$ characterizing the right moving heterotic sector is set
equal to the superperiod matrix $\hat \Omega _{IJ}$ on the left moving
superstring sector. The phases $\eta _{\delta , \bar \delta}$ and $\eta
_\delta$ are chosen so as to be consistent with modular invariance and
are present to enforce the GSO projection independently on left and right
movers, if so desired.

\vfill\eject

%%%%%%%%%%%%%%%%%%%%%%%%%%%%%%%%%%%%%%%%%%%%%%%%%%%%%%%%%%%%%%%%%%%%%
%%%%%%%%%%%%%%%%%%%%%%%%%%%%%%%%%%%%%%%%%%%%%%%%%%%%%%%%%%%%%%%%%%%%%
\section{Integrating out Odd Supermoduli}
\setcounter{equation}{0}
%%%%%%%%%%%%%%%%%%%%%%%%%%%%%%%%%%%%%%%%%%%%%%%%%%%%%%%%%%%%%%%%%%%%%
%%%%%%%%%%%%%%%%%%%%%%%%%%%%%%%%%%%%%%%%%%%%%%%%%%%%%%%%%%%%%%%%%%%%%

As a consistent formulation of the scattering amplitudes as integrals
over supermoduli space is thus available, the question now becomes as to
whether and how a consistent formulation as integrals on moduli space may
be obtained. In other words, how the odd supermoduli can be integrated
out. We shall restrict to addressing this question for genus 2 and
even spin structures, where a rigorous treatment is now available.
Comments on higher genus will be made in the last section of this paper
at the level of conjecture.

\subsection{Naive derivation of the BRST picture changing Ansatz}

It was shown in \cite{vv1} that the formulation of superstring amplitudes
of \cite{fms} may be recovered from the one in terms of supermoduli space
provided certain assumptions are made on the choices of gauge slice. Of
course, given that the work of \cite{vv1} also demonstrates that the
Ansatz of \cite{fms} leads to slice dependent formulas, it must be that
the assumptions made in the derivation of this Ansatz are incorrect.
It is very instructive to see where they fail.

\medskip

The key assumption made by \cite{vv1} in their rederivation of \cite{fms}
by integrating over the odd supermoduli is that the metric and gravitino
slices are chosen so that
\bea
\label{ass}
g_{mn} (m^a) 
\qquad \qquad
\chi = \sum _{\alpha =1,2} \zeta ^\alpha \chi _\alpha (m^a)
\eea
The interpretation of these formulas is that the bosonic moduli are
associated with the metric $g_{mn}$, independently of the odd supermoduli
$\zeta ^\alpha$. The functions $\chi _\alpha$ characterize the slice
chosen for the odd supermoduli, and are taken to be of pointlike support
at insertion points $z_\alpha$. This choice leads direcly to the Ansatz
of \cite{fms} in terms of BRST invariance and the picture changing
operator $Y(z_\alpha)$,
\bea 
\bigg \< \O \prod _{a=1}^{3h-3} (\mu _a |b) \prod _{\alpha =1} ^{2h-2}
Y (z_\alpha) \bigg \> \prod _{a=1} ^{3h-3} d m^a 
\eea
If all the assumptions had held correct, this Ansatz ought to have
produced amplitudes that are independent of the insertion points
$z_\alpha$ of the picture changing operators. An explicit calculation to
two loop level in \cite{vv1} has shown, however, that there is residual
dependence on these points.

\medskip

This situation spells disaster. If an analogy were sought with the
quantization of Yang-Mills theory, the present situation would be as if 
the perturbative evaluation of a gauge invariant correlation function of
gauge invariant operators in $\xi$-gauge were to yield a result that
is not independent of $\xi$. In Yang-Mills theory, it is clear that this
situation signals a faulty gauge fixing procedure. So it does for
superstring theory.

\subsection{The key role of local supersymmetry}

Odd supermoduli may be viewed as fibers over even supermoduli and the
operation of integrating out the odd supermoduli may be viewed as a
projection along the fibers of supermoduli space onto its even base.
The assumptions made in the previous subsection are equivalent to the
following projection,
\bea
\label{badproj}
(g_{mn}, \chi _m ) \hskip .25in & \sim & \hskip .3in (g_{mn} ', \chi _m ')
\hskip .5in {\rm under \ \ SUSY}
\nonumber \\
\downarrow \hskip .5in  & & \hskip .55in \downarrow
\nonumber \\
g_{mn} \hskip .4in & \sim \! \! \! \! \! / & \hskip .45in g_{mn} '
\hskip .75in {\rm under \ \ Diff} \times {\rm Weyl} 
\eea 
The interpretation of this diagram is as follows. The projection onto the
even moduli amounts to omitting $\chi$, as the moduli $m^a$ are functions
of only the metric and not $\chi$, cfr (\ref{ass}). The local
supersymmetry transformation of the metric is given by
\bea
\delta g_{mn}  = 2 \xi ^+ \chi _{\{m} {}^+ e_{n\}} {}^{\bar z}
\eea
This variation produces a change in the moduli $m^a$ defined above,
so that the moduli $m^a$ are defined in a manner that is not invariant
under local supersymmetry. Therefore, the projection (\ref{badproj})
itself is inconsistent with local supersymmetry and it stands to reason
that its use will lead to ambiguities.

\medskip

A consistent projection can only be obtained when the even supermoduli
$m^a$ are defined in a manner invariant under the action of local
supersymmetry. Therefore, moduli should be viewed as defined by another
metric $\hat g_{mn}$. The action of local supersymmetry
transformations on this metric $\hat g_{mn}$ must descend to an action of
diffeomorphisms and Weyl transformations only, without the admixture of
variations in moduli.  Schematically, this type of consistent
supersymmetric projection may be represented as follows,
\bea
\label{consproj}
(g_{mn}, \chi _m ) \hskip .3in & \sim & \hskip .3in (g_{mn} ', \chi _m ')
\hskip .5in {\rm under \ \ SUSY}
\nonumber \\
\downarrow \hskip .6in  & & \hskip .6in \downarrow
\nonumber \\
\hat g_{mn} (m^a) \hskip .25in & \sim & \hskip .35in 
\hat g_{mn} ' (m^a)
\hskip .5in {\rm under \ \ Diff} \times {\rm Weyl}
\eea 
The fundamental guarantee that this projection exists and has the desired
properties rests on the fact that the super period matrix $\hat \Omega
_{IJ}$, introduced earlier, is invariant under local supersymmetry. 
The form of $\hat g_{mn}$ will be needed at intermediate stages of the
calculation, but when all parts are assembled, the chiral measure will
involve only the conformal class of $\hat \Omega _{IJ}$, which is
uniquely determined by $\hat \Omega _{IJ}$. 

\medskip

Henceforth, we shall use this consistent  projection (\ref{consproj}) for
genus 2; comments on higher genus will be deferred to the last section.

\vfill\eject

%%%%%%%%%%%%%%%%%%%%%%%%%%%%%%%%%%%%%%%%%%%%%%%%%%%%%%%%%%%%%%%%%%%%%
%%%%%%%%%%%%%%%%%%%%%%%%%%%%%%%%%%%%%%%%%%%%%%%%%%%%%%%%%%%%%%%%%%%%%
\section{Construction of the Chiral Measure}
\setcounter{equation}{0}
%%%%%%%%%%%%%%%%%%%%%%%%%%%%%%%%%%%%%%%%%%%%%%%%%%%%%%%%%%%%%%%%%%%%%
%%%%%%%%%%%%%%%%%%%%%%%%%%%%%%%%%%%%%%%%%%%%%%%%%%%%%%%%%%%%%%%%%%%%%

It will be helpful to spell out the key ingredients in the construction
the chiral measure.

\medskip

$\bullet$ 
We make use of supersymmetric supermoduli $m^A = (\hat \Omega _{IJ} ,
\zeta ^\alpha)$.

\medskip

$\bullet$ 
All quantities, calculated originally for the metric $g_{mn}$ with complex
structure $\Omega _{IJ}$ are re-expressed in terms of the super period
matrix $\hat \Omega _{IJ}$. In correlation functions, this change is
achieved via the insertion of the stress tensor,
\bea
\Omega _{IJ} \to \hat \Omega _{IJ} 
\quad 
\left \{ \matrix{
g & \to & \hat g = g + \hat \mu \cr
\p _{\bar z} & \to & \hat \p _{\bar z} = \p _{\bar z} + \hat \mu \p
_z\cr
\< \cdots \> (g) & = & \< \cdots \> (\hat g) + \int \hat \mu
\< T \cdots \> (\hat g) \cr}
\right .
\eea
The Beltrami differential $\hat \mu$ is associated with the
deformation of complex structure $\Omega _{IJ}$ to $\hat \Omega _{IJ}$.
In view of the relation between the two period matrices, $\hat \mu$ is
determined by
\bea
\int _\Sigma \hat \mu \omega _I \omega _J = 
{1\over 8\pi}\int _\Sigma \! d^2 \! z \int _\Sigma
\! d^2 \! w \ \omega_I(z) \chiz  S_{\delta}(z,w) \chiw \omega_J(w)
\eea
Although ultimately only the conformal class of $\hat \mu$ will enter,
calculations at intermediate stages of the chiral measure will appeal to
the actual metric $\hat g$. The final cancellation of all dependence on
the choice of metric $\hat g$ within its conformal class
serves both as a check of the consistency of the approach,
and of the actual calculations.  

\medskip

$\bullet$ Superholomorphic forms on supermoduli space project to
holomorphic forms on moduli space (with complex structure $\hat \Omega
_{IJ}$) plus exact differentials. The simplest example is provided
by the superholomorphic 1/2 forms, which obey $\D_- \hat
\omega _I=0$ and project as follows,
\bea
\hat \omega _I (\Omega , \zeta) = \theta \omega _I (\hat \Omega ) + \D _+
\Lambda _I
\eea
The nature of this projection for general forms is of great interest for
the study of scattering amplitudes; it will be discussed in detail in the
forthcoming papers \cite{V,VI}.

\medskip

$\bullet$ The fact that the super period matrix is unchanged under
variations of the odd supermoduli $\zeta $ implies that none of the
components of the super Beltrami differential introduced in
(\ref{superbeltrami}) will vanish,
\bea
\label{complication}
\delta _\zeta \hat \Omega _{IJ} =0 \quad  
\Rightarrow \quad
\left \{ \matrix{H_A = \bar \theta (\mu _A - \theta \chi _A) \cr
\mu _A \not=0 \quad \& \quad \chi _A \not=0\cr} \right .
\eea
This situation is contrasted with that of the choice (\ref{ass})
where, for example, $\mu _\alpha=0$. 

\medskip

$\bullet$ Dual to the super Beltrami differential are superholomorphic
3/2 forms of odd type $\Phi _{IJ}$ and of even type $\Phi _\alpha$.
Of key importance will be the explicit formula for
\bea
\label{Phi}
\Phi _{IJ} = -{i\over 2} \bigg ( \hat \omega _I \D _+ \hat \omega _J + 
\hat \omega _J \D _+ \hat \omega _I \bigg )
\eea
which is normalized to satisfy $\< H_a | \Phi _{IJ}\> =
\delta _{a,IJ}$ and $\< H_\alpha | \Phi _{IJ}\> = 0$.

\vfill\eject

%%%%%%%%%%%%%%%%%%%%%%%%%%%%%%%%%%%%%%%%%%%%%%%%%%%%%%%%%%%%%%%%%%%%%
%%%%%%%%%%%%%%%%%%%%%%%%%%%%%%%%%%%%%%%%%%%%%%%%%%%%%%%%%%%%%%%%%%%%%
\section{Calculation of the Chiral Measure}
\setcounter{equation}{0}
%%%%%%%%%%%%%%%%%%%%%%%%%%%%%%%%%%%%%%%%%%%%%%%%%%%%%%%%%%%%%%%%%%%%%
%%%%%%%%%%%%%%%%%%%%%%%%%%%%%%%%%%%%%%%%%%%%%%%%%%%%%%%%%%%%%%%%%%%%%

Before starting the computation of the chiral measure, using the
ingredients developed in the preceeding sections, one more obstacle 
must be overcome.

\subsection{Change of basis of super Beltrami differentials}

First, a change of basis is to be performed on the super Beltrami
differentials. This is needed because our use of supersymmetric bosonic
moduli forces all components of $H_A$ to be non-vanishing, as indicated in
(\ref{complication}). Without it, the product of $\delta (\<H_A |B\>)$
factors would produce an exceedingly complicated and untractable form for
the correlation functions.

\medskip

Under the assumption that the vertex operators occuring in $\O$ are
independent of the ghost superfield $B$ (as is the case for all NS vertex
operators \cite{dpvertex}), the pairing of $H_A$ is effectively with a
superholomorphic $B$ field. Therefore, we may change basis from
$H_A$ to new super Beltrami differentials $H_A ^*$, 
chosen for maximal simplicity to be
\bea
H^* _a & = & \bar \theta \delta (z,p_a) \hskip 1in a=1,2,3
\nonumber \\
H^* _\alpha & = & \bar \theta \theta \delta (z,q_\alpha)
\hskip .9in \alpha =1,2
\eea
Denoting an arbitrary complete set of linearly independent even and
odd superholomorphic 3/2 forms by $\Phi _C$, we have 
\bea
\prod _A \delta (\<H_A | B \> ) 
= {\sdet \<H_A | \Phi _C \> \over \sdet \< H_A ^* | \Phi _C \> }
\prod _ a b(p_a) \prod _\alpha \delta (\beta (q_\alpha))
\eea
Clearly, this formula is independent of the choice of $\Phi _C$. Its
considerable advantage is that all correlation functions are now with
respect to standard insertions and all the complication inherent in $H_A$
has been relegated to a single multiplicative factor.

\medskip

Two natural choices of basis emerge for $\Phi_C$. The first, denoted
simply by $\Phi _C$, is dual to $H_A$, while the second, denoted by $\Phi
^* _C$, is dual to $H^* _A$,
\bea
\< H_A | \Phi _C \> = \< H_A ^* | \Phi ^* _C \> = \delta _{AC}
\eea
The explicit form of the basis vectors $\Phi ^* _C$ is known and may be
found in \cite{II}, Appendix B. The explicit form for $\Phi _C$ is only
known for half of its components. Indeed, it was already established in
(\ref{Phi}) that $\Phi _{IJ}$ provides the odd components of $\Phi _C$,
namely for the even indices $c = \{IJ\}$. The even partners $\Phi
_\gamma$, however, have no such canonical expression. On general grounds,
they may be expressed as a linear combination,
\bea
\Phi _\gamma (\z) = \Phi _\epsilon ^* (\z) C^\epsilon {} _\gamma + 
\Phi _{IJ} (\z) D^{IJ} {}_\gamma
\eea
for some  $\z$-independent, but moduli dependent, matices $C$ and $D$. 
By pairing with $H_\alpha$ and using the fact that $\< H_\alpha |\Phi
_{IJ}\>=0$, we readily have $\det C \times \det \<H_\alpha |\Phi ^*
_\gamma \>=1$. Taking all factors into account, 
\bea
{\sdet \<H_A | \Phi _C \> \over \sdet \< H_A ^* | \Phi _C \> }
=
{1 \over \det \Phi _{IJ} (p_a) \times \det \<H_\alpha |\Phi ^* _\gamma\>}
\eea
The components $\Phi _{IJ} (p_a)$ are explicitly known. The components
$\mu _\alpha$ and $\chi _\alpha$ of $H_\alpha = \bar \theta (\mu _\alpha -
\theta \chi_\alpha)$ are known in the following manner. The objects
$\chi_\alpha$ represent the choice of worldsheet gravitini slice, and
should be viewed as input into the gauge fixing process (which, in the
end is to be independent of the choice of $\chi _\alpha$). The object
$\mu _\alpha$ may be shown \cite{II} to be given by $\mu _\alpha = \p \hat
\mu / \p \zeta ^\alpha$. All ingredients in the gauge fixed formula are
thus known explicitly, and we have the following formula for the chiral
measure,
\bea
\A [\delta] 
=
{ \< \prod _a b(p_a) \prod _\alpha \delta (\beta (q_\alpha)) \> 
\over 
\det \Phi _{IJ+} (p_a) \det \< H_\alpha | \Phi _\beta ^* \> }
\biggl \{ 1 + {1 \over 2 \pi} \int \hat \mu \< T \>
- {1 \over 8 \pi ^2} \int _\Sigma \! \! \int _\Sigma \chi \chi \< S S \> 
\biggr \}  
\eea
As a consistency check, it was demonstrated in \cite{II} that this
expression is indeed invariant under local supersymmetry on the
worldsheet, as is expected on general grounds.

\subsection{The calculation in components}

In order to achieve workable formulas, the above expression is henceforth
considered for a gravitino slice supported at two arbitrary generic points
$x_1$ and $x_2$,
\bea
\chi _\alpha (z) = \delta (z,x_\alpha)
\eea
The chiral measure may then be expressed entirely in terms of quantities
that are meromorphic on the worldsheet. These include the prime form
$E(z,w)$, the Szeg\"o kernel $S_\delta (z,w)$, the $b-c$ ghost Green
function $G_2(z,w)$ (which is defined to vanish when $z=p_1,p_2,p_3$ in
view of the $b$-insertions at $p_a$) and the superghost Green function
$G_{3/2}(z,w)$ (which is defined to vanish when $z=q_1,q_2$ in view of
the $\delta (\beta )$-insertions at $q_\alpha$). They also include  a
number of holomorphic differentials;  $\psi _\alpha ^* (z)$ and
$\bar \psi _\alpha (z)$ are holomorhic 3/2 forms normalized so that $\psi
_\alpha ^* (q_\beta ) =\bar \psi _\alpha (x_\beta )= \delta _{\alpha
\beta}$, and the quantity $\varpi _a (z,w)$ provides a one-to-one map
between holomorphic 2 forms in one variable and holomorphic forms of two
variables of weight 1 each. It obeys the normalization $ \varpi _a (p_b,
p_b) =\delta _{ab}$. 

\medskip

The chiral measure is given as follows,
\bea
{\cal A} [\delta] 
=
{ \< \prod _a b(p_a) \prod _\alpha \delta (\beta (q_\alpha))  \>  \over 
\det \omega _I \omega _J (p_a)  \cdot \det  \psi ^* _\beta  (x_\alpha)}
\biggl \{ 1  + {\zeta ^1 \zeta ^2 \over 16 \pi ^2} \sum _{i=1}^6 
{\cal X}_i \biggr \} 
\eea
with the following expressions for $\X_i$, 
\bea
\label{bigg}
\X_1  & = & 
 -10 S_\delta (x_1,x_2) \p _{x_1} \p _{x_2} \ln E(x_1,x_2)
\nonumber \\
&&
-3 \p _{x_2} G_2 (x_1, x_2) G_{3/2}(x_2,x_1) 
- 2 G_2 (x_1,x_2) \p _{x_2} G_{3/2}(x_2,x_1)  -(1 \leftrightarrow 2)
\nonumber \\ && \nonumber \\
\X _2  &=& 
S_\delta (x_1,x_2) \omega _I(x_1) \omega _J(x_2)  
\p _I \p _J \ln \biggl ({\tet [\delta ](0)^5 \tet (p_1+p_2+p_3-3\Delta )
\over
\tet [\delta ](q_1+q_2-2\Delta )} \biggr )
\nonumber \\ && \nonumber \\
\X _3  &=& 
2 S_\delta (x_1,x_2) \sum _a  \varpi _a  (x_1, x_2) \bigl [B_2(p_a)
+ B_{3/2}(p_a) \bigr ]
\nonumber \\ && \nonumber \\
\X_4  &=&  2
S_\delta (x_1,x_2) \sum _a 
\p _{p_a} \p _{x_1} \ln E(p_a,x_1) \varpi  _a(p_a, x_2)
- (1 \leftrightarrow 2)
 \\ && \nonumber \\
\X _5  &=& 
\sum _a  
S_\delta (p_a, x_1) \p _{p_a} S_\delta (p_a,x_2)  
\varpi _a  (x_1,x_2) -(1 \leftrightarrow 2) 
\nonumber \\ && \nonumber \\
\X_6  &=& 
3  \p _{x_2} G_2(x_1,x_2) G_{3/2} (x_2,x_1)
+ 2 f_{3/2} (x_1) G_2(x_1,x_2) \p \bar \psi _1 (x_2)
    -(1 \leftrightarrow 2) 
\nonumber \\
&&  
+ 2 G_{3/2} (x_2,x_1)  G_2(x_1,x_2) \p \bar \psi _2(x_2) 
+ \p _{x_2} G_2(x_2,x_1) \p \bar \psi _2 (x_1)
    -(1 \leftrightarrow 2)
\nonumber
\eea
where we have used the following notations, 
\bea
f_n(w) & = & \omega_I(w)\p_I{\rm ln}\,\tet [\delta](D_n)
+\p_w{\rm ln} (\prod_i \sigma (w) E(w,z_i) ) 
\nonumber \\
B_2(w) & = & -27 T_1(w) +
 \half f_2 (w)^2 - {3 \over 2} \p _w f_2 (w)
-2 \sum _a \p_{p_a} \p _w \ln E(p_a,w) \varpi _a (p_a, w) 
\nonumber \\
B_{3/2} (w) & = & 12 T_1(w) - \half f_{3/2}(w)^2 + \p _w f_{3/2}(w) 
+ {3 \over 2} \p _{x_1} G_2 (w,x_1) + {3 \over 2} \p _{x_2} G_2 (w,x_2) 
\nonumber \\
&& 
- {3 \over 2} \p _w G_{3/2}(x_1,w) \bar \psi _1 (w)
- {3 \over 2} \p _w G_{3/2}(x_2,w) \bar \psi _2 (w)
- {1 \over 2}  G_{3/2}(x_1,w) \p \bar \psi _1 (w)
\nonumber \\
&&
- {1 \over 2}  G_{3/2}(x_2,w) \p \bar \psi _2 (w)
+ G_2 (w,x_1) \p \bar \psi _1 (x_1) + G_2 (w,x_2) \p \bar \psi _2 (x_2)
\eea

\subsection{Fundamental Consistency Check} 

The above expression for the chiral measure is a sum of terms that are
manifestly well-defined scalar meromorphic functions of $x_\alpha$,
$q_\alpha$ and $p_a$. By inspection of any possible singularities when a
given point approaches any of the remaining 6 points, it may be shown
directly \cite{III} that the above result is actually holomorphic in each
point, and thus independent of all 7 points
$x_\alpha$, $q_\alpha$ and $p_a$. This important result checks that our
approach and calculations are indeed consistent.

\subsection{Corrections to picture changing operators}  

In view of the independence on the points $x_\alpha$, $q_\alpha$ and
$p_a$, it is very interesting to reinvestigate the problems that emerged
in the old approach of \cite{fms} and \cite{vv1}. In particular, in the
old approach, the product of the picture changing operators was singular.
In our approach, no such singularity can emerge. Their cancellation
proceeds as follows. The only contribution in (\ref{bigg}) common with
\cite{fms} and \cite{vv1} arises from $\X_1$ which contains
\bea
\X_1 \sim \< S(x_1)  \delta (\beta (q_1)) \ S(x_2) \delta (\beta (q_2))
\ b(p_1) b(p_2) b(p_3) \> 
\eea
Formally, the picture changing operator is defined by $Y(q) \sim S(q)
\delta (\beta (q))$ in the limit  $x_\alpha \to q_\alpha$. However, this
limit on the terms in $\X_1$ alone,  
\bea
G_{3/2}(x_1,x_2)  \to 
\left \{ \matrix{  0 & {\rm as} & x_1 \to q_1 \cr
\infty & {\rm as} & x_2 \to q_2} \right .  
\eea
is ill-defined and singular. An identical and opposite singularity arises
from the finite-dimensional determinants summarized in $\X_6$, however,
and the combination of both contributions is regular. Thus, a correct
definition of the picture changing operators must include the
appropriate contributions of associated finite-dimensional determinants.

\subsection{The choice of a convenient gauge}

\begin{itemize}

\item
Given the independence of the chiral measure on all points, we
may set $x_\alpha = q_\alpha$. 

\item
Furthermore, terms $\X_2$, $\X_3$, $\X_4$ are proportional to $S_\delta
(x_1,x_2) = S_\delta (q_1,q_2)$ and will  vansih upon choosing the {\sl
split gauge} $S_\delta (q_1,q_2)=0$. This gauge is also natural since it
implies  $\hat \Omega _{IJ} = \Omega _{IJ}$. 

\item
Finally, it is advantageous to choose the points $p_a$ to be the three
zeros of a holomorphic 3/2 form $\psi _A (z)$. This choice yields a
particularly useful form for the
$b-c$ Green function $G_2$ in terms of
$\psi _A$ and the Szego kernel,
\bea
G_2 (z,w) = S_\delta (z,w) \psi_A (z)/\psi_A (w) 
\eea
so that in split gauge we have $G_2 (q_1, q_2)=0$. Combining all
contributions, we now have $\X_1+\X_6=\X_2= \X_3 = \X_4=0$, and only $\X
_5 \not= 0$ remains.

\end{itemize}

The evaluation of the single remaining term $\X_5$ is quite involved and
will not be reproduced here; it may be found in \cite{IV}. The final
result of the calculation  will be discussed in the next section.

\vfill\eject

%%%%%%%%%%%%%%%%%%%%%%%%%%%%%%%%%%%%%%%%%%%%%%%%%%%%%%%%%%%%%%%%%%%%%
%%%%%%%%%%%%%%%%%%%%%%%%%%%%%%%%%%%%%%%%%%%%%%%%%%%%%%%%%%%%%%%%%%%%%
\section{Explicit Formulas in terms of $\tet$-constants}
\setcounter{equation}{0}
%%%%%%%%%%%%%%%%%%%%%%%%%%%%%%%%%%%%%%%%%%%%%%%%%%%%%%%%%%%%%%%%%%%%%
%%%%%%%%%%%%%%%%%%%%%%%%%%%%%%%%%%%%%%%%%%%%%%%%%%%%%%%%%%%%%%%%%%%%%

The fundamental result of \cite{IV} is a concise formula for the
chiral superstring measure in terms of $\tet$-constants and modular
forms. Henceforth, only the superperiod matrix will appear, which we
shall now denote by $\Omega _{IJ}$ to simplify notation. The measure is,
\bea
d \mu [\delta] (\Omega ) =   {\Xi _6 [\delta ](\Omega ) \ \tet 
[\delta ]^4  (0,\Omega) \over
16 \pi ^6 \ \Psi _{10} (\Omega) }  \  d^3 \Omega _{IJ} \
\eea
It remains to explain the various ingredients in this formula.

\medskip

$\bullet $
On a genus 2 surface, there are 16 independent spin structures,
which may be labelled by half integer characteristics,
\bea
\kappa = (\kappa ' | \kappa '')
\qquad \qquad 
\kappa ', \kappa '' \in (0,\half)^2
\eea
Here the two components $\kappa '_I$ of $\kappa '$ refer to the spin
structure assignments along the homology cycles $A_I$, while the
components $\kappa ''_I$ refer to those on the cycles $B_I$. 
One distinguishes even and odd spin structures according to whether
$4\kappa ' \cdot \kappa ''$ is even or odd. We have,
\bea
\cases{
\kappa  & even/odd iff $4 \kappa ' \cdot \kappa ''$ even/odd \cr
\delta  & 10 even spin structures \cr
\nu     & \  6 odd spin structures\cr
\delta  & = $\nu _{i_1}+\nu _{i_2}+\nu _{i_3} 
= \nu _{i_4}+\nu _{i_5}+\nu _{i_6}$ \cr}
\eea
The last lines states the fact, specific to genus 2, that every even spin
structure may be written (in exactly two different ways) as the sum of
three distinct odd spin structures. A compact notation will be used
for the pairing,
\bea
\< \kappa | \rho \> \equiv \exp \{ 4 \pi i (\kappa ' \cdot \rho '' -
\rho ' \cdot \kappa '')\}
\eea
which by construction takes on the values $\pm 1$.

\medskip

$\bullet$
The $\tet$-functions with characteristic $\kappa$ are defined by
\bea
\tet [\kappa](v, \Omega) \equiv \sum _{n - \kappa ' \in {\bf Z}^2}
\exp \bigl \{ i \pi n^t \Omega n + 2 \pi i n^t (v+ \kappa '')\bigr \}
\eea
and are manifestly holomorphic in all arguments. They are even/odd
functions of $v \in {\bf C}^2$ according to whether $\kappa$ is an
even/odd spin structure.

\medskip

$\bullet$ 
We shall make heavy use of modular forms for genus 2. These were
classified long ago in \cite{igusa}. Using the $\tet$-constants $\tet
[\delta](0,\Omega)$, one may readily produce an infinite series of
modular forms, for $k=1,2,3,\cdots$
\bea
\Psi _{4k} (\Omega ) \equiv \sum _\delta \tet [\delta ] ^{8k} (0,\Omega )
\eea
These forms are not all independent. Instead, they form a polynomial
ring with a finite number of generators \cite{igusa}. We shall need in
particular the form of weight 10 given by,
\bea
\Psi _{10} (\Omega ) \equiv \prod _{\delta \ {\rm even}} 
\tet [\delta ]^2 (0, \Omega)
\eea

\medskip

$\bullet$ Finally, the evaluation introduces a new modular quantity, $\Xi
_6 [\delta ](\Omega)$ which may be defined as follows. Let the even spin
structure $\delta$ be decomposed as the sum of three distinct odd spin
structures $\delta = \nu_1 + \nu_2 + \nu_3=\nu_4+\nu_5+\nu_6$. We then
have,
\bea
\label{xiodd}
\Xi _6 [\delta](\Omega) \equiv \sum _{1 \leq i< j \leq 3} \!
\<\nu_i | \nu_j\> \! \prod _{k=4,5,6} \! \tet [\nu_i + \nu _j 
+\nu_k]^4 (0,\Omega)
\eea
Alternatively, $\Xi_6[\delta]$ may be expressed via even spin
structures only, 
\bea
\label{xieven}
\Xi _6 [\delta](\Omega) =  \sum _{[\delta, \delta _1,
\delta_2, \delta _3]} \biggl ( - \half \prod _{i=1} ^3
\<\delta | \delta _i \>  \tet [\delta _i ]^4 (0,\Omega)
\biggr )
\eea
To speficy the sum over the even spin structures $\delta _i$, $i=1,2,3$,
we introduce the symbol $e$ as well as the following nomenclature
\cite{igusa},
\bea
e (\delta, \epsilon, \eta) \equiv \<\delta | \epsilon \>
\<\epsilon | \eta \>\< \eta | \delta \> =
\cases{
+ 1 &  syzygous \ triple \cr
- 1 &  asyzygous \ triple \cr }
\eea
The sum over quartets $[\delta, \delta _1, \delta_2, \delta _3]$ is
defined to be such that any of its 4 distinct triplets is asyzygous.
\bea
\left \{ \matrix{
e(\delta_1 , \delta _2, \delta _3) = e(\delta , \delta _1, \delta _2)
& = & -1  \cr 
e(\delta , \delta _2, \delta _3) = e(\delta , \delta _3, \delta
_1) & = & -1 \cr} \right . 
\eea 
Note that while the definition of $\Xi_6 [\delta]$ given in (\ref{xiodd})
is restricted to genus 2, the form given in (\ref{xieven}) makes sense
for any genus, and may be viewed as a definition of $\Xi_6[\delta]$ in
arbitrary genus. Also note that $\Xi _6 [\delta]$ is not a modular form,
since it has an explicit dependence on the spin structure $\delta$. There
does exist a  modular form $\Psi _6$ of weight 6, obtained by summing 
products of three $\tet ^4$ over all 60 asyzygous (or 60 syzygous)
triplets in the following formula,
\bea
\Psi _6 (\Omega ) \equiv \sum _{e(\delta_1,\delta _2,\delta _3)=-1}
\pm \tet [\delta _1]^4 (0,\Omega) \tet [\delta _2]^4 (0,\Omega) \tet
[\delta _3]^4 (0,\Omega)
\eea 
But this modular form is not the same object as $\Xi_6[\delta]$.

\vfill\eject

%%%%%%%%%%%%%%%%%%%%%%%%%%%%%%%%%%%%%%%%%%%%%%%%%%%%%%%%%%%%%%%%%%%%%
%%%%%%%%%%%%%%%%%%%%%%%%%%%%%%%%%%%%%%%%%%%%%%%%%%%%%%%%%%%%%%%%%%%%%
\section{Modular Properties -- GSO Phases}
\setcounter{equation}{0}
%%%%%%%%%%%%%%%%%%%%%%%%%%%%%%%%%%%%%%%%%%%%%%%%%%%%%%%%%%%%%%%%%%%%%
%%%%%%%%%%%%%%%%%%%%%%%%%%%%%%%%%%%%%%%%%%%%%%%%%%%%%%%%%%%%%%%%%%%%%

Modular transformations are defined to leave the canonical
intersection matrix invariant, and thus form the group $Sp(4,{\bf Z})$,
\bea
\left ( \matrix{A & B \cr C & D \cr} \right ) 
\left ( \matrix{0 & I \cr -I & 0 \cr} \right )
\left ( \matrix{A & B \cr C & D \cr} \right ) ^t
=
\left ( \matrix{0 & I \cr -I & 0 \cr} \right )
\qquad \qquad
\left ( \matrix{A & B \cr C & D \cr} \right ) \in Sp(4,{\bf Z})
\eea
The action on spin structure  is given by \cite{igusa}
\bea
\left (\matrix{ \tilde \kappa' \cr \tilde \kappa ''\cr}  \right )
=
\left ( \matrix{D & -C \cr -B & A \cr} \right )
\left ( \matrix{ \kappa ' \cr \kappa '' \cr} \right )
+ \half \ {\rm diag} 
\left ( \matrix{CD^T  \cr AB^T \cr} \right )\, ,
\eea
Here, diag$(M)$ of an $n\times n$ matrix $M$ is an $1\times n$ column
vector whose entries are the diagonal entries of $M$.
On the period matrix, the transformation acts by
\bea
\tilde \Omega = (A\Omega + B ) (C\Omega + D)^{-1}
\eea
while on the Jacobi $\tet$-functions, the action is given by
\bea
\tet [\tilde \kappa ] \biggl ( \{(C\Omega +D)^{-1} \}^t v , \tilde
\Omega \biggr ) =
\epsilon (\kappa, M) \det (C\Omega + D) ^\half 
e^{ i \pi v ^t (C\Omega +D)^{-1} C v }
\tet [ \kappa ] (v, \Omega)
\eea
The phase factor $\epsilon (\kappa, M)$ depends upon both $\kappa $ and
the modular transformation $M$ and obeys  $\epsilon (\kappa , M )^8=1$.
As a result, we obtain the following transformation laws
\bea
d^3 \tilde\Omega _{IJ}
& = & \det (C\Omega+D)^{-3} \ d^3 \Omega_{IJ}
\nonumber \\
\tet [\tilde \delta ] ^4 (0,\tilde \Omega) & = & 
\epsilon ^4 \ \det (C\Omega + D)^2 \ \tet [\delta ] ^4 (0,\Omega)
\nonumber \\
\Xi _6 [\tilde \delta ] (\tilde \Omega) 
& = & 
\epsilon ^4 \ \det (C\Omega + D)^6 \ \Xi _6 [\delta ] (\Omega )
\nonumber \\
\Psi _{10} (\tilde \Omega) 
& = & \det (C\Omega + D)^{10} \ \Psi _{10}(\Omega)
\eea
Note that $\Xi_6[\delta ](\Omega)$ does not transform as a
modular form.

\medskip

The modular transformation law of the chiral measure may be readily
deduced, 
\bea
\label{modmeasure}
d\mu [\tilde \delta] (\tilde \Omega) = 
\det\,(C\Omega+D)^{-5}d\mu [\delta ](\Omega)
\eea
The weight $-5$ is related to the critical dimension, 10, as
may be seen most easily after the integration over internal momenta has
been carried out. The resulting factor of $\det \Im \Omega$ has the
following modular transformation law,
\bea
\det \Im \tilde \Omega = |\det (C\Omega +D)|^{-2} \det \Im \Omega
\eea 
Therefore, the full measure combining left and right movers is
modular covariant,
\bea
(\det \Im \tilde \Omega)^{-5} \ d\mu [\tilde \delta] (\tilde \Omega)
\times \overline{d\mu [\tilde {\bar \delta}] (\tilde \Omega)}
=
(\det \Im \Omega)^{-5} \ d\mu [ \delta] ( \Omega) \times \overline{d\mu [
{\bar \delta}] ( \Omega)} 
\eea
The phase factor in (\ref{modmeasure}) is 1 for every spin structure, so
that all GSO phases, consistent with modular invariance, must be equal
and may be set to 1. Notice that this phase assignment is unique.

\vfill\eject

%%%%%%%%%%%%%%%%%%%%%%%%%%%%%%%%%%%%%%%%%%%%%%%%%%%%%%%%%%%%%%%%%%%%%
%%%%%%%%%%%%%%%%%%%%%%%%%%%%%%%%%%%%%%%%%%%%%%%%%%%%%%%%%%%%%%%%%%%%%
\section{Behavior under Degenerations}
\setcounter{equation}{0}
%%%%%%%%%%%%%%%%%%%%%%%%%%%%%%%%%%%%%%%%%%%%%%%%%%%%%%%%%%%%%%%%%%%%%
%%%%%%%%%%%%%%%%%%%%%%%%%%%%%%%%%%%%%%%%%%%%%%%%%%%%%%%%%%%%%%%%%%%%%

A key consistency check on the calculation of any superstring measure or
scattering amplitude is that it must obey the proper factorizations onto
physical states when the string worldsheet degenerates. The general
structure of such factorizations is known on general physical grounds.
This check was carried out in \cite{IV}. There are two inequivalent 
cases, according to whether the degeneration separates the
surface into two disconnected parts or not. We make the choice of
canonical homology cycles of Fig.\ref{fig:2}, and use the following
parametrization of the period matrices in this homology basis,
\bea
\Omega = \left ( \matrix{\tau_1 & \tau \cr \tau & \tau _2} \right )
\eea
Clearly, the {\sl separating degeneration} corresponds to the limit $\tau
\to 0$ as $\tau_1$ and $\tau _2$ are kept fixed, while the non-separating
degeneration may be taken to correspond to the limit $q \equiv \exp \{ i
\pi \tau_2\} \to 0$ as $\tau _1 $ and $\tau$ are kept fixed. The two
cases are illustrated in Fig.\ref{fig:3}.

%%%%%%%%%%%%%%%%%%%%%%

\begin{fig}[htp]
\centering
\epsfxsize=6in
\epsfysize=4.3in
\epsffile{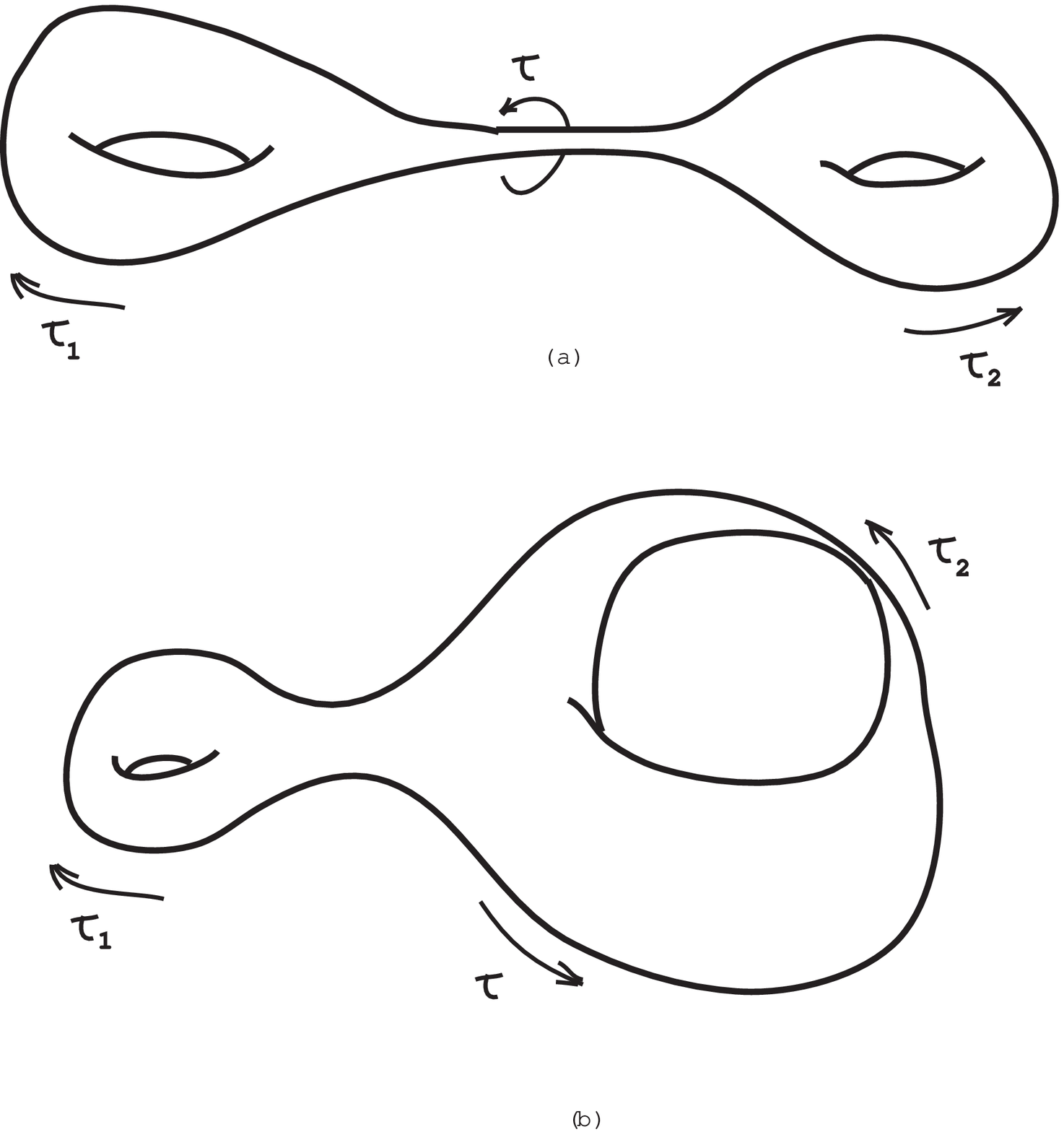}
\caption{Degenerations of a genus 2 surface : (a) separating, (b)
non-separating.}
\label{fig:3}
\end{fig}

%%%%%%%%%%%%%%%%%%%%%

\subsection{Separating Degeneration : $\tau \to 0$, $\tau_{1,2}$ fixed}

In this limit, we distinguish between two cases. First is the case of 9
out of the 10 even spin structures $\delta$ for which the genus 1 spin
structures $\mu_1$ and $\mu_2$ on the genus 1 connected components are
both even; this is the NS-NS case. Second is the case of the single even
spin structure for which the genus 1 spin structures are both odd and
equal to $\nu_0$; this is the R-R case. The asymptotic behavior in both
cases may be worked out using the limit of the $\tet$-function,
which may be expressed in the following way,
\bea
\label{thetaasym}
\tet \left [ \matrix{\mu _1 \cr \mu _2 \cr} \right ](0,\Omega )
&=&
\sum _{p=0} ^\infty {(2 \tau) ^{2p} \over (2p)!} 
\p _{\tau _1}^p \tet _1[\mu _1] (0,\tau _1) 
\p _{\tau _2}^p \tet _1[\mu _2] (0,\tau _2) 
\nonumber \\
\tet \left [ \matrix{\nu_0 \cr \nu_0 \cr} \right ](0,\Omega )
&=&
{1 \over 4 \pi i} \sum _{p=0} ^\infty {(2 \tau )^{2p+1} \over (2p+1)!} 
\p _{\tau _1}^p \tet _1 '[\nu_0] (0,\tau _1) 
\p _{\tau _2}^p \tet _1 '[\nu_0] (0,\tau _2) 
\eea
Here, $\tet_1$ are  genus 1 $\tet$-functions. The limits of $\Psi
_{10}(\Omega)$ and $\Xi_6[\delta](\Omega)$ are given by 
\bea
\label{xiasym}
\Psi _{10} (\Omega ) & = & 
- (2 \pi \tau )^2 \cdot 2^{12} \cdot \eta (\tau _1)^{24} \eta (\tau
_2)^{24}  +\O(\tau ^4)
\nonumber \\
\Xi _6 \left [ \matrix{\mu _1 \cr \mu _2 \cr} \right ] (\Omega)
& = &
- 2^8 \cdot \<\mu_1 |\nu_0\> \<\mu _2 |\nu_0\> \eta (\tau _1)^{12} \eta
(\tau _2)^{12}  +\O(\tau ^2)
\nonumber \\
\Xi _6 \left [ \matrix{\nu _0 \cr \nu _0  \cr} \right ] (\Omega)
& = & 
-3 \cdot 2^8 \cdot \eta (\tau _1)^{12} \eta (\tau _2)^{12}  +\O(\tau ^2)
\eea
Combining all contributions, the limit of the measure is found to be
\bea
{\rm NS-NS } \hskip .8in
d\mu \left [ \matrix{  \mu_1 \cr \mu_2  \cr} \right ]
& = &
{ d^3 \tau \over \tau ^2} 
\prod _{i=1,2}   {\< \mu _i |\nu _0\> \tet  [\mu_i ]^4 (\tau_i)
\over  32 \pi ^4 \eta (\tau _i)^{12} } + \O(\tau^0)
\nonumber \\
{\rm R-R } \hskip 1in
d\mu \left [ \matrix{ \nu_0 \cr \nu_0  \cr} \right ]
& = &
{ 3 \tau ^2 d^3 \tau \over 2^6 \pi ^4} + \O (\tau ^4)
\eea 
The NS-NS case reproduces the correct 1-loop factors, including the GSO
phases appropriate for 1-loop amplitudes. Notice that these phases
emerged from the limit of $\Xi_6[\delta]$. The $\tau ^{-2}$ prefactor
indicates the presence of the tachyon intermediate state, which is indeed
expected in the NS-NS sector.  Upon combining this limiting measure with
the right moving part, and including the effects of the internal momenta,
a massless pole will also be present. Partial GSO summation in one or the
other 1-loop component will cancel both the tachyon and the massless
singularities, as is expected. The R-R case has neither tachyon nor
massless intermediate singularities, as indeed is expected.

\subsection{Non-Separating Degeneration : $q \to 0$, $\tau_1,\tau$
fixed}

The case of non-separating degenerations is analogous, and we only quote
the results here; detailed derivations may be found in \cite{IV}. We
have
\bea
d\mu \left [ \matrix{\mu_i \cr 00  \cr} \right ]   
& = & + \
V_i (\tau, \tau_1) {d^3 \tau \over q}  + \O(q^0)
\nonumber \\
d\mu \left [ \matrix{\mu_i \cr 0\half  \cr} \right ]   
& = & - \
V_i (\tau, \tau_1) {d^3 \tau \over q}  +
\O(q^0)
\nonumber \\
d\mu \left [ \matrix{\mu_i \cr \half 0  \cr} \right ]  & = & \O(q^0)
\nonumber \\
d\mu \left [ \matrix{\nu_0 \cr \nu_0  \cr} \right ] & = & \O(q^0)
\eea
In the first three lines, $\mu_i$ stands for any of the three even genus
1 spin structures on handle 1, and $V_i(\tau, \tau_1)$ stands for the
tachyon 2-point function on the degenerate genus 1 surface. The $q^{-1}$
singularity corresponds to the tachyon traversing the homology cycle
$A_2$ when the spin structure in handle 2 is either $[00]$ or $[0 \ 1/2]$,
which corresponds to NS boundary conditions. This is as expected. Note
that a partial summation over spin structures in the NS sector alone
eliminates the tachyon singularity, again as expected. On the other hand,
no tachyon appears when the spin structures $[1/2 \ 0]$ and $[1/2 \ 1/2]$
correspond to R boundary conditions, again as expected. 

\medskip

To conclude, the measure passes all immediate degeneration checks carried
out in this section, confirming its validity.

\vfill\eject

%%%%%%%%%%%%%%%%%%%%%%%%%%%%%%%%%%%%%%%%%%%%%%%%%%%%%%%%%%%%%%%%%%%%%
%%%%%%%%%%%%%%%%%%%%%%%%%%%%%%%%%%%%%%%%%%%%%%%%%%%%%%%%%%%%%%%%%%%%%
\section{Vanishing of the Cosmological Constant}
\setcounter{equation}{0}
%%%%%%%%%%%%%%%%%%%%%%%%%%%%%%%%%%%%%%%%%%%%%%%%%%%%%%%%%%%%%%%%%%%%%
%%%%%%%%%%%%%%%%%%%%%%%%%%%%%%%%%%%%%%%%%%%%%%%%%%%%%%%%%%%%%%%%%%%%%

General arguments have been given long ago that space-time supersymmetry
guarantees the vanishing of the cosmological constant in superstring
theory considered in flat space-time \cite{martinec}. This vanishing is
just one  example of a whole array of non-renormalization results in
supersymmetric Yang-Mills theory and string theory (for recent reviews see
\cite{dp99,magoo,df02}). As discussed in the Introduction, attempts
had been made by several authors \cite{vv1,mhns,lp,ars2} to derive this
vanishing from a first principles calculation, but, as explained earlier,
progress was halted by the ambiguities that were believed to plague
superstring perturbation theory. Now that we have a consistent formula for
the measure available, the vanishing of the cosmological constant may be
derived from a first principles calculation in superstring theory.

\medskip

The two-loop contribution to the cosmological constant for both the Type
II and Heterotic superstrings are the most immediate quantities that
may be evaluated once the superstring chiral measure is known. 
They are given by
\bea
\Lambda _{{\rm II}} & = & 
\int _{\M_2} {|d^3 \Omega|^2 \over  (\det \Im \Omega)^5 } \times
 {\Upsilon _8 (\Omega ) \overline{\Upsilon _8 (\Omega )} \over 2^8 \pi
^{12} |\Psi _{10} (\Omega)|^2}
\nonumber \\
\Lambda _{{\rm H}} & = & 
\int _{\M_2} {|d^3 \Omega|^2 \over  (\det \Im \Omega)^5 } \times
{\Upsilon _8 (\Omega ) \overline{\Psi _8 (\Omega )} \over 2^8 \pi
^{12} |\Psi _{10} (\Omega)|^2}
\eea
We have used the fact that the chiral measure for the
26-dimensional bosonic string, partially compactified on the Cartan tori
of $E_8 \times E_8$ or Spin$(32)/{\bf Z}_2$ is proportional to $\Psi
_8(\Omega )/\Psi _{10} (\Omega)$, as was established in \cite{IV}.
 
\medskip

The remaining ingredient is the chiral measure for the left moving
superstring, which is obtained by summing over all even spin structures of
\bea
\Upsilon _8 (\Omega ) 
\equiv 
\sum _\delta \Xi_6 [\delta ](\Omega ) \tet [\delta ]^4 (0,\Omega ) 
\eea
Here, we have made use of the unique GSO phase factor assignment that
follows from requiring modular invariance. 

\medskip

We shall now show that $\Upsilon _8 =0$, so that both cosmological
constants vanish $\Lambda _{{\rm II}}= \Lambda _{{\rm H}}=0$. First,
$\Upsilon _8$ is a modular form of weight 8, by its very construction.
Next, we know from the asymptotics derived in (\ref{thetaasym}) and
(\ref{xiasym}) that $\Upsilon _8 \to 0$ in the limit of separating
degeneration. By the general classification of genus 2 modular forms of
\cite{igusa}, any modular form of weight 8 must be proportional to $\Psi
_4 (\Omega )^2$, so that we must have $\Upsilon _8 (\Omega ) = c\Psi _4
(\Omega )^2$ for some (moduli independent) constant
$c$. But $\Psi _4 (\Omega)$ tends to a non-zero value in the limit of
separating degeneration. Therefore, the constant $c$ and thus
$\Upsilon _8 (\Omega)$ must vanish. Note that this relation amongst
$\tet$-constants implied by the vanishing of $\Upsilon_8$ does not result
from the Riemann relations alone and is equivalent instead to the
relation $\psi _4 (\Omega )^2 = 4 \Psi _8 (\Omega)$.

\vfill\eject

%%%%%%%%%%%%%%%%%%%%%%%%%%%%%%%%%%%%%%%%%%%%%%%%%%%%%%%%%%%%%%%%%%%%%
%%%%%%%%%%%%%%%%%%%%%%%%%%%%%%%%%%%%%%%%%%%%%%%%%%%%%%%%%%%%%%%%%%%%%
\section{Scattering Amplitudes}
\setcounter{equation}{0}
%%%%%%%%%%%%%%%%%%%%%%%%%%%%%%%%%%%%%%%%%%%%%%%%%%%%%%%%%%%%%%%%%%%%%
%%%%%%%%%%%%%%%%%%%%%%%%%%%%%%%%%%%%%%%%%%%%%%%%%%%%%%%%%%%%%%%%%%%%%

The vertex operators for the scattering of $N$ massless bosons
are given by
\be
\prod_{i=1}^N V(k_i,\epsilon_i)
= \prod_{i=1}^N \int _\Sigma \! d^{2|2} \z_i\, E(\z_i)\,
\epsilon_i ^{\mu _i} \bar \epsilon_i ^ {\bar \mu _i}
{\cal D}_+X^{\mu_i} {\cal D}_-X^{\bar\mu_i}
e^{ik_i \cdot X} (\z_i)
\ee
As in the case of the measure, the superstring scattering amplitudes
require a GSO summation over spin structures of the conformal blocks of 
$\<\prod_{i=1}^NV(k_i,\epsilon_i)\>_X$ in the $X^{\mu}$ superconformal
field theory.

\subsection{Vanishing of the 1-, 2- and 3-point functions}

On general grounds, the vanishing of the 1-, 2- and 3-point functions is
expected from space-time supersymmetry \cite{martinec} and, using our
measure, may be shown from first principles. The 0-, 1-, 2- and
3-point functions in both the Type II and the heterotic strings are then
found to vanish pointwise on moduli space and without the appearance of
boundary terms, in view of the following new identities, 
\bea
&& 
\sum _\delta \Xi _6 [\delta] (\Omega)  \tet [\delta ] (0,\Omega)^4
S_\delta (z_1,z_2) ^2 =0
\nonumber \\
&& 
\sum _\delta \Xi _6 [\delta] (\Omega)  \tet [\delta ] (0,\Omega) ^4 
S_\delta(z_1,z_2) S_\delta (z_2,z_3) S_\delta (z_3,z_1) =0
\eea
which were proven in \cite{IV} using the Fay trisecant identity
\cite{fay,vv2}. A full discussion of the calculations and proofs
involved will be presented in a forthcoming paper \cite{V}.

\subsection{The 4-point function}

The 4-point function receives contributions from two distinct parts. The
first arises from the connected part of the correlators
\bea
\label{connected}
\< S(z) S(w) \prod_{i=1}^4 V(k_i,\epsilon_i)^{chi} \>_{\rm conn}
\qquad {\rm and} \qquad
\< T(z) \prod_{i=1}^4 V(k_i,\epsilon_i)^{chi} \>_{\rm conn}
\eea
The second arises from the disconnected part
\bea
\label{disconnected}
\< S(z) S(w) \> \< \prod_{i=1}^4 V(k_i,\epsilon_i)^{chi} \>
\qquad {\rm and} \qquad
\< T(z)\> \< \prod_{i=1}^4 V(k_i,\epsilon_i)^{chi} \>
\eea
of these correlators and combines with the gauge fixing determinants into
a contribution proportional to the measure $d\mu [\delta](\Omega)$. The
connected part is more complicated and requires an independent treatment
to appear in a later publication \cite{VI}. 

\medskip

The disconnected part (for example for the Type II superstrings) is given
by
\bea
\label{fourpoint}
\<\prod_{i=1}^4V(\epsilon_i,k_i)\>
=
g_s ^2 \delta (k) \int _{\M_2} 
{\bigl | d\Omega ^3 \bigr |^2 \over  (\det \,\Im \Omega)^5} 
\prod_{i=1}^4 \int _\Sigma d^2z_i\, \bigl | {\cal F} \bigr |^2 
\exp \biggl (-\sum_{i<j}k_i \! \cdot \! k_jG(z_i,z_j)\biggr )
\eea
Here, $g_s$ is the string coupling, the scalar Green's function is given
by
\be
G(z,w) = -{\rm log}|E(z,w)|^2 + 2\pi \Im \int_z^w \!
\omega_I \ (\Im\,\Omega)^{-1}_{IJ} \Im \int_z^w \! \omega_J
\ee 
while $k$ is the total momentum, and $\F$ is a holomorphic 1-form in
each $z_i$, given by
\bea
\F = C_{{\cal S}}\, {\cal S} (1234) + \sum _{(i, j,k) = {\rm perm}(2,3,4)}
C_{{\cal T}} (1i|jk) \,{\cal T}(1i|jk) 
\eea
The combinations $C_{{\cal S}}$ and $C_{{\cal T}}$ are kinematical
factors, which depend only on the polarization vectors $\epsilon _i$ and
the external momenta $k_i$ through the gauge invariant combinations
$f_i ^{\mu \nu} \equiv \epsilon _i ^\mu k _i ^\nu - \epsilon _i ^\nu k_i
^\mu$ and are given by
\bea
C_{{\cal S}} &=&
f_1 ^{\mu \nu} f_2 ^{\nu \mu} f_3 ^{\rho \sigma} f_4 ^{\sigma \rho} +
f_1 ^{\mu \nu} f_2 ^{\rho \sigma} f_3 ^{\nu \mu} f_4 ^{\sigma \rho} +  
f_1 ^{\mu \nu} f_2 ^{\rho \sigma} f_3 ^{\sigma \rho} f_4 ^{\nu \mu}
\\ &&
- 4 f_1 ^{\mu \nu} f_2 ^{\nu \rho} f_3 ^{\rho \sigma} f_4 ^{\sigma \mu} 
- 4 f_1 ^{\mu \nu} f_2 ^{\rho \sigma} f_3 ^{\nu \rho} f_4 ^{\sigma \mu} 
- 4 f_1 ^{\mu \nu} f_2 ^{\nu \rho} f_3 ^{\sigma \mu} f_4 ^{\rho \sigma} 
\nonumber \\
C_{{\cal T}} (ij|kl) &=&
f_i ^{\mu \nu} f_j ^{\rho \sigma} f_k ^{\nu \mu} f_l ^{\sigma \rho} -
f_i ^{\mu \nu} f_j ^{\rho \sigma} f_k ^{\sigma \rho} f_l ^{\nu \mu} +2
f_i ^{\mu \nu} f_j ^{\nu \sigma} f_k ^{\sigma \rho} f_l ^{\rho \mu} -2
f_i ^{\mu \nu} f_j ^{\nu \sigma} f_k ^{\rho \mu} f_l ^{\sigma \rho} 
\nonumber 
\eea
The kinematical combination $C_{{\cal S}}$ coincides with the unique
kinematical invariant of the NS 4-point function encountered at tree and
1-loop level, which is often expressed in terms of the rank 8 tensor $t$
(see \cite{gs82, grosswitten}), $C_{\cal S} = - 8 t _{\kappa _1 \lambda
_1 \kappa _2 \lambda _2 \kappa _3 \lambda _3 \kappa _4 \lambda _4} f _1
^{\kappa _1 \lambda _1} f_2 ^{\kappa _2 \lambda _2} f _3 ^{\kappa _3
\lambda _3} f _4 ^{\kappa _4 \lambda _4}$.
Finally, the forms ${\cal S}$ and ${\cal T}$ are given by
\bea
{\cal S} (1234) &=& 
- {1 \over 192 \pi ^6 \Psi _{10} } \ 
\omega _I (z_1) \omega _J (z_2) \omega _K (z_3) \omega _L(z_4) 
\sum _\delta \Xi _6 [\delta ] \tet [\delta ]^3 \p_I \p_J \p_K \p_L \tet
[\delta](\Omega)
\nonumber \\
{\cal T} (ij|kl) &=&
- {1 \over 8 \pi ^2} \ \omega _{[1} (z_1) \omega _{2]} (z_2) \omega _{[1}
(z_3) \omega _{2]} (z_4)
\eea
The $\delta$-sum for the ${\cal T}$-term was carried out explicitly, and
no $\Psi _{10}$ appears in its contribution. ${\cal S}$ and $C_{{\cal S}}$
are totally symmetric, while ${\cal T}$ and $C_{{\cal T}}$ are odd under
the interchange of $i\leftrightarrow j$ or $k\leftrightarrow l$. As a
result, the ${\cal T}$-term is novel at 2 loops and could not exist at 1
loop.

\subsection{Finiteness of the disconnected part}

The disconnected part of the $4$-point function for massless bosons,
calculated above, is finite. This is the case at least when the
Mandelstam variables $k_i\cdot k_j$ are purely imaginary. As is now well
known \cite{dp94}, finiteness for general $k_i\cdot k_j$
cannot be read off directly, but has to be established by analytic
continuation.

\medskip

To show convergence, we recall that the modular form $\Psi_{10}(\Omega)$
vanishes of second order along the divisor of separating nodes. This
corresponds to the propagation of a tachyon, and was responsible for the
divergence in the bosonic string \cite{div}. Here, the modular
tensor
\bea
\label{modtensor}
\sum_{\delta}
\Xi_6[\delta](\Omega)\tet ^3[\delta](\Omega)
\p_I \p_J \p_K \p_L \tet [\delta](\Omega)
\eea
also vanishes of second order along the divisor of
separating nodes, rendering the superstring amplitude finite.
We illustrate this cancellation mechanism in a typical case,
where, say, two of the indices $I,J,K,L$ are with respect
to the variable $\zeta_1$ in $\tet[\delta](\zeta,\Omega)$,
and the other two are with respect to the variable $\zeta_2$.
In this case, the asymptotics  are given by
\bea
\label{pthetaasym}
\p_I \p_J \p_K \p_L \tet \left [ \matrix{\mu _1 \cr \mu _2 \cr} 
\right ](0,\Omega )
&=&
-2^4\pi^2 
\p _{\tau _1}^p \tet _1''[\mu _1] (0,\tau _1) 
\p _{\tau _2}^p \tet _1''[\mu _2] (0,\tau _2) 
+{\cal O}(\tau^2)
\nonumber \\
\p_I \p_J \p_K \p_L 
\tet \left [ \matrix{\nu_0 \cr \nu_0 \cr} \right ](0,\Omega )
&=&
2^5\pi^3i\,\tau\,\p_{\tau_1}\eta^3(\tau_1)
\p_{\tau_2}\eta^3(\tau_2)+{\cal O}(\tau^3)
\eea
It follows from (\ref{thetaasym})
that the asymptotics of the $\tet$-constants themselves
are given by
\bea
\label{thetaasym1}
\tet \left [ \matrix{\mu _1 \cr \mu _2 \cr} \right ](0,\Omega )
&=&
\tet _1[\mu _1] (0,\tau _1) 
\tet _1[\mu _2] (0,\tau _2)+{\cal O}(\tau^2) 
\nonumber \\
\tet \left [ \matrix{\nu_0 \cr \nu_0 \cr} \right ](0,\Omega )
&=&
{1 \over 2 \pi i}\,\tau\,  
\p _{\tau _1} \tet _1 '[\nu_0] (0,\tau _1) 
\p _{\tau _2} \tet _1 '[\nu_0] (0,\tau _2)+{\cal O}(\tau^3) 
\eea
Combining these factors in the disconnected part of the 4-point function,
we see that the contribution of the last spin structure in
(\ref{thetaasym1}) is of order ${\cal O}(\tau^4)$, and can be ignored.
The sum over the remaining even spin structures $\delta$ produces, up to
${\cal O}(\tau^2)$
\bea
&&
\sum_{\delta}
\Xi_6[\delta](\Omega)\tet ^3[\delta](\Omega)
\p_I \p_J \p_K \p_L \tet [\delta](\Omega)
=
2^{12}\pi^2\eta^{12}(\tau_1)\eta^{12}(\tau_2)
\\
&&\quad\times\,\sum_{\mu_1}\langle \mu_1|\nu_0\rangle
\tet_1^3[\mu_1](0,\tau_1)
\p_{\tau_1}\tet_1[\mu_1](0,\tau_1)
\,\sum_{\mu_2}\langle \mu_2|\nu_0\rangle
\tet_1^3[\mu_2](0,\tau_2)\p_{\tau_2}\tet_1[\mu_2](0,\tau_2)
\nonumber
\eea
This vanishes, in view of the derivative of the Jacobi identity.
Next, we discuss the case of non-separating degenerations. As explained
earlier in section \S 10.2, the degenerating parameter is then $q\to 0$.
The asymptotics of $\Psi_{10}(\Omega)$  are given by
\be
\Psi_{10}(\Omega)
=
-2^{12}\,q^2\,\eta^{18}(\tau_1)\tet^2[\nu_0](\tau,\tau_1)
+
{\cal O}(q^3)
\ee
so the finiteness will result from the vanishing to order ${\cal O}(q^2)$
of (\ref{modtensor}). To see that this is indeed the case, we note that
for all even genus 2 spin structures $\delta$ whose component along
cycle 2 produces an R sector along the degenerating $B_2$ cycle, 
both the terms $ \tet[\delta](0,\Omega)$ and $\p_I\p_J\p_K\p_L
\tet[\delta](0,\Omega)$ vanish to order $\O (q^{1\over 4})$, while
$\Xi_6[\delta](\Omega)$ vanishes to order $\O (q)$.
Thus the contributions of all such spin structures are of
order ${\cal O}(q^2)$, and can be ignored in the proof of finiteness.
The remaining genus 2 even spin structures $\delta$ correspond to the
NS sector along the $B_2$ cycle and are of the form
\bea
\delta _{{\rm NS}} = \left [ \matrix{\mu \cr (0|\kappa'') \cr} \right ]
\eea
The asymptotics of $\tet [\delta _{{\rm NS}}]$  and 
$\p_I\p_J\p_K\p_L\tet[\delta _{{\rm NS}}](0,\Omega)$ 
are, up to order $\O(q^2)$, invariant under the interchange
$\kappa''=0 \leftrightarrow \kappa''= 1/2$.
On the other hand, under the same interchange, the
asymptotics of $\Xi_6 [\delta _{{\rm NS}}] $
are odd, again up to order ${\cal O}(q^2)$.
It follows that the contribution to (\ref{modtensor}) of the spin
structures $\delta_{NS}$ vanishes up to the order ${\cal O}(q^2)$.
We observe that, as expected on physical grounds, 
the cancellation mechanism here does not
require the one-loop Jacobi identity.

\subsection{The Supergravity Limit}

In the low energy limit, the exponential factor of the scalar Green's
function in (\ref{fourpoint}) tends to 1. It is instructive to identify
the kinematical factors that emerge from the integration over the 4
vertex insertion points $z_i$ of $| \F |^2$ in (\ref{fourpoint}) in the
Type II superstrings (analogous expressions may be derived for the
heterotic strings). The first contribution is from the product
$C_{\cal S} \bar C _{\cal S}$, and yields the well-known $t t R^4$ term of
four Riemann tensors contracted with two copies of the rank 8 tensor $t$
 as obtained in \cite{grosswitten}. As argued in the
preceding paragraph, this contribution is given by a convergent integral.
The second contribution is from the products $C_{\cal S} \bar C_{\cal
T}$; it vanishes in view of the complete symmetry in the points $z_i$ in
${\cal S}$ and the antisymmetry in two pairs of points in ${\cal T}$. The
third contribution is from the product $C_{\cal T} \bar C_{\cal T}$ for
which the $z_i$ integrals may be carried out using the Riemann
bilinear relations. The resulting kinematical factors is again a
quadrilinear in the Riemann tensor and is proportional to
\bea
C_{\cal T} \bar C_{\cal T} & \longrightarrow &
+  \bigl (
R_{\alpha \beta \mu \nu} R^{\alpha \beta \mu \nu} \bigr )^2 
-  
R_{\alpha \beta \mu \nu} R^{\gamma \delta \mu \nu}
R^{\alpha \beta \rho \sigma } R_{\gamma \delta \rho \sigma}
\nonumber \\ &&
+ 4 
R_{\alpha \beta \mu \nu} R^{\gamma \delta \mu \nu} 
R^{\beta}{}_{ \gamma \rho \sigma} R_{\delta}{}^{ \alpha \rho \sigma}
- 4 
R^{\alpha \beta \mu \nu} R_{\delta \alpha \mu \nu} 
R_{\beta \gamma \rho \sigma} R^{\gamma \delta \rho \sigma}
\nonumber \\ &&
+ 4 
R^{\alpha \beta \mu \nu} R_{\beta \gamma \nu \rho}
R^{\gamma \delta \rho \sigma} R_{\delta \alpha \sigma \mu}
- 4 
R^{\alpha \beta \mu \nu} R_{\beta \gamma \nu \rho}
R_{\delta \alpha}{}^{ \rho \sigma} R^{\gamma \delta}{}_{ \sigma \mu}
\eea
While it is possible that this term, which arose from the
{\sl disconnected contributions} in (\ref{disconnected}), will be
cancelled by similar contributions arising from the {\sl connected
contributions} in (\ref{connected}), the above contribution to the low
energy effective action has at least one remarkable property~: the
integral over moduli space becomes simply the volume of moduli space with
respect to the $Sp(4,{\bf Z})$ invariant volume form $|d^3\Omega |^2
(\det \Im \Omega )^{-3}$. We note that the problem of loop corrections in
Type II superstrings and their contribution to low energy effective
actions has witnessed a resurgence of interest recently (see for example
\cite{greenvanhove,opiz,wester}).

\vfill\eject

%%%%%%%%%%%%%%%%%%%%%%%%%%%%%%%%%%%%%%%%%%%%%%%%%%%%%%%%%%%%%%%%%%%%%
%%%%%%%%%%%%%%%%%%%%%%%%%%%%%%%%%%%%%%%%%%%%%%%%%%%%%%%%%%%%%%%%%%%%%
\section{Compactification \& The Cosmological Constant  }
\setcounter{equation}{0}
%%%%%%%%%%%%%%%%%%%%%%%%%%%%%%%%%%%%%%%%%%%%%%%%%%%%%%%%%%%%%%%%%%%%%
%%%%%%%%%%%%%%%%%%%%%%%%%%%%%%%%%%%%%%%%%%%%%%%%%%%%%%%%%%%%%%%%%%%%%

One of the most important paradoxes in theoretical physics is the
extraordinary smallness of the cosmological constant as compared to
typical particle physics scales. In theories with global space-time
supersymmetry, such as super Yang-Mills theories, general arguments show
that the cosmological constant must be comparable to the scale at which
supersymmetry is broken. In theories with local space-time supersymmetry,
such as supergravity and superstring theory, the situation is different
and it is possible to have vanishing cosmological constant despite the
fact that supersymmetry is broken \cite{ferrara}.

\subsection{Motivation from Orbifold Compactifications}

Recently, this alternative has been investigated in a number of
papers \cite{ks,ksiz} within the context of orbifold compactifications.
The models of \cite{ks} are constructed in a manner such that the
cosmological constant vanishes to 1-loop level, despite the fact that
supersymmetry is broken; it was proposed that the cosmological constant
should vanish also to higher orders. Lacking reliable formulas for
multiloop amplitudes, however, it was impossible to check these claims
with concrete calculations. We shall now revisit these issues.
 
\medskip

In orbifold compactifications with non-Abelian orbifold groups,
qualitatively novel effects emerge starting at two loop level. The
reason for this is as follows.  The orbifolding of one of the
10-dimensional flat space-time superstring theories is carried out by
making identifications of the string fields $x^\mu$ and $\psi ^\mu$ under
an orbifold group $G$, which is a subgroup of the Poincar\'e group
$ISO(1,9)$,
\bea
\rho : \pi _1(\Sigma) \to G \qquad \left \{ \matrix{
x (\gamma z) = \rho _x  (\gamma) \ x (z)
\cr
\psi (\gamma z) = \rho _\psi  (\gamma) \ \psi  (z)
\cr } \right .
\eea
as the point $z$ is taken around a homotopy cycle $\gamma \in \pi _1
(\Sigma)$ on $\Sigma$. The maps $\rho _x (\gamma)$ and $\rho _\psi
(\gamma)$ are representations of $\pi _1 (\Sigma)$ in $G$. The
representations  $\rho _x$ and $\rho_\psi$ will in general be different
because $x$ may be translated as well as rotated, while $\psi$ may only
be rotated.

\medskip

Some of the most interesting cases are when $G$ is non-Abelian \cite{ks}.
To tree and one loop levels, however, $\pi _1 (\Sigma)$ is Abelian and
therefore runs through an Abelian subgroup of $G$. Thus, one-loop
contributions to the cosmological constant in non-Abelian orbifold models
can capture only part of the orbifold phenomena. To two loop level and
higher, $\pi _1 (\Sigma)$ is always non-Abelian and the non-Abelian
effects of the orbifolding process will start to play a role.

\subsection{The Cosmological constant for general compactifications}

The calculations of the cosmological constant presented earlier for flat
space-time may be extended to the case where some of the space-time
directions are compactified to form a total space-time manifold $C$,
under the following mild assumptions \cite{I,adp}.

\medskip

$\bullet $ The compactification only modifies the matter conformal field
theory, leaving the superghost part unchanged;

$\bullet$ The compactification respects $\N=1$ local worldsheet
supersymmetry, so that the super-Virasoro algebra with matter central
charge $c=15$ is preserved.

\medskip

Chiral splitting must be carried out with some care, as the superconformal
families will be labeled no longer only by the internal loop momenta
$p^\mu _I$ of flat, characteristic of flat space-time. Instead,
superconformal families will be labeled by $f$, which may include internal
loop momenta as well as twist sectors and any other quantum numbers
specifying the superconformal families. The spin structure $\delta$ will
not be included in $f$ as this label also enters into the ghost
superconformal field theory. We shall denote by $Z_M [\delta] (\Omega)$
the partition function for flat Minkowski space-time (omitting the
Gaussian factor involving the internal momenta), and by
$Z_C [\delta ] (\Omega , f)$ the partition function for compactification
onto the manifold $C$, associated with superconformal family $f$.

\medskip

Under these assumptions, the superstring measure is independent of any
choices of gauge slice, and a simple expression was derived in \cite{I}
for split gauge $S_\delta (q_1,q_2)=0$,
\bea
d \mu _C [\delta] (\Omega, f) 
=   {Z_C [\delta] (\Omega, f) \over Z_M [\delta ](\Omega)}
\biggl \{ {\Xi _6 [\delta ]  \tet [\delta ]^4 \over 4 \Psi _{10}} (\Omega)
- {\cal Z} \< S_C (q_1) S_C (q_2)  \>_C (\Omega , f)
\biggr \} {d^3 \Omega  \over 4 \pi ^6}
\qquad
\eea
where
\bea
{\cal Z} \equiv {\< \prod _a b(p_a) \prod _\alpha \delta  (\beta
(q_\alpha)) \>
\over \det \omega _I \omega _J (p_a) }
\qquad \qquad
Z_M [\delta ](\Omega) \equiv {\tet [\delta ]^5 \over Z^{15}}
\eea
Here, $S_C$ denotes the supercurrent of the compactified theory, and $Z$
is the chiral boson partition function. The actual cosmological constant
is obtained by assembling contributions from left and right movers. For
Type II theories, we have
\bea
\Lambda _{{\rm II}} & = & \int _{\M_2} \sum _{\delta, \bar \delta} \sum
_{f,\bar f} M(\delta, f;\bar \delta, \bar f) \times d\mu _C [\delta
](\Omega, f) \times d\mu _C [\bar \delta ](\bar \Omega, \bar f)
\eea
and an analogous formula may be derived for the Heterotic strings. The
matrix $M(\delta, f; \bar \delta, \bar f)$ represents a Hermitian metric
on the space of superconformal blocks and must be chosen consistently with
modular invariance. Implementation of these results on specific orbifold
compactification models will be deferred to \cite{adp}.

\vfill\eject

%%%%%%%%%%%%%%%%%%%%%%%%%%%%%%%%%%%%%%%%%%%%%%%%%%%%%%%%%%%%%%%%%%%%%
%%%%%%%%%%%%%%%%%%%%%%%%%%%%%%%%%%%%%%%%%%%%%%%%%%%%%%%%%%%%%%%%%%%%%
\section{Comments on Higher Loops}
\setcounter{equation}{0}
%%%%%%%%%%%%%%%%%%%%%%%%%%%%%%%%%%%%%%%%%%%%%%%%%%%%%%%%%%%%%%%%%%%%%
%%%%%%%%%%%%%%%%%%%%%%%%%%%%%%%%%%%%%%%%%%%%%%%%%%%%%%%%%%%%%%%%%%%%%

The case of genus 2 is of particular importance because it is the lowest
order where odd supermoduli play a non-trivial role. The case of
genus higher than 2 is expected to be significantly more difficult,
since a number of simplifying features, special to genus 2,
will be then absent. Reliable explicit calculations are not
available at the present time, and we shall limit our discussion
to a few speculative remarks.

\medskip

The most encouraging fact for string perturbation theory in higher
genus is that the chiral splitting procedure of \cite{dp88, dp89}
holds for any genus. In particular, the super period matrix exists for any
genus, and is given by
\bea
\hat\Omega_{IJ}
=
\Omega_{IJ}-{i\over 8\pi}\int _\Sigma \! d^2\! z \int _\Sigma \! d^2\! w
\ \omega_I(z) \chiz  \hat S_{\delta}(z,w) \chiw \omega_J(w)
\eea
where the modified Szego kernel $\hat S_\delta (z,w)$ is defined
recursively by the relation
\bea
\p _{\bar z} \hat S_\delta (z,w)
+ {1 \over 8 \pi} \chiz \int _\sigma d^2 x \p_z \p_x \ln E(z,x)
\chi _{\bar x} {}^+ \hat S_\delta (x,w) = 2 \pi \delta (z,w)
\eea
The key projection onto the super period matrix is thus well-defined
for any genus.

\medskip

In genus 3 and higher, new technical and conceptual difficulties arise.
First, the parametrization of the fiber of the projection onto the super
period matrix will be more complicated. Second, the Dirac operator may
develop zero modes even for even spin structures. For example, in genus
3, the number of zero modes jumps from 0 to 2 on the complex codimension
1 subvariety of genus 3 hyperelliptic surfaces. Therefore, the relation
between the period matrix $\Omega _{IJ}$ and the super period matrix $\hat
\Omega_{IJ}$ may become singular at hyperelliptic
Riemann surfaces.\footnote{The importance of this fact was stressed to us
by Edward Witten.} We expect however that the full chiral string measure
will remain smooth at this locus, because the fermion determinant factor
produces multiple zeros precisely at the same locus. Actually, a related
mechanism is known to occur for the bosonic string measure in higher
genus. While the bosonic string measure is known to be singularity-free
inside moduli space \cite{dp86}, (see also \cite{moore}) it has apparent
singularities when expressed in terms of $\tet$-functions and modular
forms. In fact, its denominator in such an expression is the product
$\Psi_{18}(\Omega)$ of all $36$ even $\tet$-characteristics, which
vanishes along the locus of hyperelliptic curves \cite{belavin}. But, as
can be anticipated on general ground, the corresponding poles are
cancelled by the remaining factors in the measure.

\medskip

Another difficulty is the issue of Schottky relations. Naively, it appears
that the gauge-fixing procedure requires both $\Omega_{IJ}$ and
$\hat\Omega_{IJ}$ to be the period matrices of two-dimensional bosonic
geometries. While such a requirement may seem difficult to satisfy in
general, we note that here, the deformation from $\hat\Omega_{IJ}$ to
$\Omega_{IJ}$ consists purely of soul elements. At least for genus 3,
outside of a lower dimensional subvariety, any positive definite
symmetric matrix $\hat\Omega_{IJ}$ is the period matrix of a Riemann
surface. One may therefore expect that the gauge fixing procedure
presented here will extend at least to the case of genus 3.

\vskip 1in

\noindent
{\large \bf Acknowledgements}

\medskip

The authors would like to thank Professors Kefeng Liu and Shing-Tung
Yau for their kind invitation to participate in  the International
Conference on String Theory in Beijing, and in the opening of
the new Mathematical Sciences Center in Hang Zhou.
They gratefully acknowledge  the opportunity given to them to lecture on
this material, and the exceptional hospitality extended 
to them during their visit by Professors Kefeng Liu, Hu Sen, Shing-Tung
Yau and Chuan-Jie Zhu. They are also happy to acknowledge several 
stimulating exchanges with Edward Witten.

\vfill\eject

%%%%%%%%%%%%%%%%%%%%%%%%%%%%%%%%%%%%%%%%%%%%%%%%%%%%%%%%%%%%%%%%%%%%%
%%%%%%%%%%%%%%%%%%%%%%%%%%%%%%%%%%%%%%%%%%%%%%%%%%%%%%%%%%%%%%%%%%%%%

\end{document}